\documentclass[11pt, a4paper,preprintnumbers]{article}

\usepackage{jheppub}
\usepackage{package-notes}
\usepackage{amsmath}
\usepackage{graphicx,amsmath,amssymb}
\usepackage{subcaption}
\usepackage{caption}
\usepackage{tabularx}
\usepackage{float}
\preprint{LAPTH-047/25}

\title{Regge trajectories for UV completions of graviton scattering  from polynomial boundedness}

\author[a]{Christopher Eckner,} 
\author[b]{Felipe Figueroa,}
\author[c]{Simon Metayer,}
\author[c]{Piotr Tourkine}

\affiliation[a]{Center for Astrophysics and Cosmology, University of Nova Gorica, Vipavska 11c, 5270 Ajdov\v{s}\v{c}ina,
Slovenia}
\affiliation[b]{Service de Physique de l’Univers, Champs et Gravitation, Université de Mons, 20 place du Parc, 7000 Mons, Belgium}
\affiliation[c]{LAPTh, CNRS et Universit\'{e} Savoie Mont-Blanc 
9 Chemin de Bellevue, F-74941 Annecy, France}

\abstract{We study graviton scattering amplitudes. Assuming they are UV completed by a theory of weakly coupled massive higher spins, we demonstrate that the UV completion must possess infinitely many Regge trajectories, and thus they are forced to have a stringy spectrum.
We extend and simplify a previous proof by some of us for open-string like states to the case of external gravitons. In the present new proof, we trace the need for infinitely many trajectories to the constraint of polynomial boundedness, ultimately tied to causality.
We further present numerical results based on the stringy ansatz of Häring-Zhiboedov, which illustrates how single-trajectory-like solutions actually emerge as extremal solutions of numerical bootstrap. In our numerics, these trajectories curiously show up as numerically very large \textit{sister} trajectories. We provide solid evidence that the solutions are spurious as they appear to admit a divergent limit for infinite ansatz size. 
}

\begin{document}
\maketitle

\section{Introduction}

In the tree-level -- weak coupling -- approximation, it is known that gravitational scattering amplitudes must look stringy~\cite{Camanho:2014apa,Caron-Huot:2016icg}. The presence of higher derivative corrections to the Einstein action induces causality violations at the macroscopic level~\cite{Camanho:2014apa} and requires that infinite towers of massive higher-spins are exchanged to restore causality. In addition, under some further assumptions\footnote{Studied recently in~\cite{Haring:2023zwu}.} such theories should possess the asymptotics of the Veneziano amplitude~\cite{Caron-Huot:2016icg}.

In this paper, we investigate the question of ``stringiness'' from a different but related angle. We start from these gravitational amplitudes with an infinite exchange of massive higher spins in the tree-level approximation.
In these amplitudes, the spectrum organizes itself in terms of Regge trajectories that connect the spins and masses of the excitations. For instance, in the Virasoro-Shapiro four graviton string amplitude we have
\begin{equation}
    J = \alpha' m_n^2,\quad m_n^2 = (2 n-2)/\alpha',\quad n\in \mathbb{N}
\end{equation}
In this context, one can ask: what is the simplest such amplitude for four-graviton scattering? Arguably, the Virasoro-Shapiro amplitude is particularly simple, with its closed form expression~\eqref{eq:VirShap} given by a single ratio of Gamma functions:
\begin{equation}
\label{eq:VirShap}
   \mathcal{M}^{++--}_{VS}(s,t,u)=- \left([12]\langle34\rangle \right)^4\frac{\Gamma(-s)\Gamma(-t)\Gamma(-u)}{\Gamma(1+s)\Gamma(1+t)\Gamma(1+u)},
\end{equation}
where the spinor prefactor simply encodes the graviton polarization tensors, in this case in the maximally helicity violating ($++--$) configuration.
The spectrum of the Virasoro--Shapiro amplitude is made, as is well known, of infinitely many parallel Regge trajectories. In contrast to this amplitude, a simpler \textit{spectrum} could instead be made of a \textit{single trajectory} of states -- with exactly one spin per mass level. 

What would be the stringiness of such an object? At a given vibrational energy, usual strings exhibit a large degeneracy of modes, whereas this simpler one would carry exactly one mode. Therefore, it would look more like a classical rotating string solution tuned to have $J=\frac1{2\pi\alpha'} M^2$.
It is the subleading Regge trajectories that encode the additional vibrational modes of the string worldsheet, which, in string theory, eventually build up the full tower of spinning degrees of freedom of a quantized relativistic string. 

In a previous paper~\cite{Eckner:2024pqt}, we proved that such simple spectra cannot be realized in open-string-like scattering, and rather infinitely many Regge trajectories are required for a tree-level stringy amplitude to be consistent with unitarity, crossing and analyticity. Our motivation stemmed from our earlier study~\cite{Eckner:2024ggx}, where we proposed a numerical scheme to describe general Regge spectra by decomposing amplitudes not as a plain sum over exchanged states, but rather as a sum over Regge trajectories themselves. Formally, such a sum cannot be truncated, though at finite numerical precision it provides an efficient approximation scheme, which is powerful enough to single out string theory as a special solution.

Several arguments with varying degrees of rigor have been brought forward in the 1960s and early 1970s that rebutted the existence of single-trajectory meromorphic amplitudes, reviewed in detail in \cite{Eckner:2024pqt}. Some work~\cite{mandula-slansky-singletraj, atkinson_conditions_1968, PhysRevLett.21.383, Ademollo:1968cno, Fujisaki:1970ce} used finite-energy sum rules, the weak point being the difficulty of proving their applicability, uniformly across the entire range of complex energies. On the other side, if the external states have different masses, it can be seen~\cite{Freedman:1966zec, Nachtmann:2003ik} that spurious poles arise at $t=0$ which can only be canceled in the presence of infinitely many daughter trajectories. A more recent argument was also formulated in \cite{Cheung:2025tbr}.
In the case of Conformal Field Theories the higher amount of symmetry gives a better handle on the problem, and a result equivalent to the requirement of infinitely many trajectories in this setup has been known for more than a decade. More precisely, lightcone bootstrap techniques show that in dimensions $d \ge 2$ there are infinitely many families of double-trace operators whose trajectories asymptotically approach lines of constant twist \cite{Komargodski:2012ek,Fitzpatrick:2012yx}. This statement has been proven rigorously in recent work \cite{Pal:2022vqc}, to which we refer the reader for a more detailed treatment of this issue in the context of CFTs.

Beyond the question of the number of Regge trajectories, several recent works have studied with analytical and numerical tools the possible spectra of weakly coupled UV completions both in the gravitational and non-gravitational cases~\cite{Albert:2022oes,Albert:2023jtd,Albert:2023seb,Albert:2024yap,Ma:2023vgc,Dong:2024omo,Saha:2024qpt,Bhat:2024agd,Bhat:2025zex,Berman:2024owc,Berman:2024eid,Cheung:2022mkw,Cheung:2023adk,Cheung:2023uwn,Cheung:2024obl,Cheung:2024uhn,Cheung:2025tbr,Huang:2025icl,Calisto:2025tjo,Komatsu:2025onf,Komatsu:2025sqo}, as well as on bootstrap techniques applied to gravitational interactions in general~\cite{Guerrieri:2021ivu,Guerrieri:2022sod,Haring:2024wyz,Bellazzini:2016xrt,Correia:2025enx,Beadle:2025cdx,Bellazzini:2025shd}.

In the present paper, we extend and simplify the theorem of~\cite{Eckner:2024pqt} to gravitational scattering in $d=4$. We study $2\to2$ graviton scattering and demonstrate that standard physical axioms imply that consistent ultraviolet completions must involve infinitely many Regge trajectories. We tie the physical meaning of the no-go to a clearer conceptual origin. The proof ultimately reduces to showing that a meromorphic amplitude with only finitely many trajectories necessarily violates a basic S-matrix axiom: polynomial boundedness, which is closely tied to causality, unitarity, and locality. Beyond the technicalities associated with the kinematics of couplings between external spinning states (gravitons) and internal higher-spin exchanges, the argument is rather simple and makes manifest the tension between polynomial boundedness and a spectrum consisting of finitely many Regge trajectories. Relaxing this requirement, we show that it is straightforward to construct amplitudes with a single trajectory, which are, however, exponentially growing.

From the point of view of classical reasoning above, it is perhaps not surprising that our theorems show a tension with a form of causality. Classical, rigid bodies are incompatible with relativity and therefore they need to possess some elasticity and vibrational degrees of freedom  \textit{at some energy}. Therefore it is also not a surprise that the theorem is not limited to gravity only, but to the relativistic UV completions of any theory involving extended objects.

This result leads to several practical questions. How small can the trajectories be? Can they start arbitrarily far in the UV? To address some of these questions, we provide in the second part of the paper a numerical primal (i.e. constructive) bootstrap based on the stringy ansatz introduced in~\cite{Haring:2023zwu}. This also allows us to make contact with recent numerical findings in the literature. With this ansatz, we study the space of allowed couplings between gravitons and higher-spin states. Both for gravitational scattering in 10 dimensions~\cite{Albert:2024yap,Huang:2025icl} and for large-N gauge theories in 4 dimension~\cite{Albert:2022oes,Albert:2023seb,Albert:2023jtd}, it was observed with a different numerical method, known as a dual bootstrap, that the spectra of extremal amplitudes appear to possess a single Regge trajectory.

As we scan the space of allowed couplings by extremizing them numerically, we observe that our resulting extremal amplitudes are ``single-trajectory-like'', in a way that we characterize precisely. The most prominent feature is the emergence of a \textit{sister} trajectory\footnote{Sister trajectories have $1/2$ the slope of the leading one, or in general $1/p$ with $p$ integer.} that dominates by many orders of magnitude the scattering amplitude, and below which states seem to decouple fully from the spectrum up to a certain mass threshold that increases with the size of the ansatz\footnote{Up to a possible {\it niece} (daughter of the sister) trajectory for which our numerics is not powerful enough (see also appendix \ref{app:ampl-gig}).}. Beyond this threshold, all subleading trajectories kick in, showing the presence of infinitely many trajectories at sufficiently high energies. States \textit{above} the emergent trajectory are basically left untouched and we speculate on the relation between our numerics and some elements that could have appeared to be numerical artefacts in the aforementioned papers.

While the scale at which trajectories below the emergent sister trajectory are activated is pushed to infinity with growing ansatz size, we also observe that this limit is singular, because the amplitude itself becomes infinite. 
We speculate that this might signal something else. As the amplitudes become exponentially large compared to the gravitational term (even at finite $s,t$), naively this would imply that the amplitude cannot remain weakly coupled.
Therefore, if we insist on maintaining weak coupling, the subleading trajectories should be visible, and the string and all of its excitations should be fully resolvable. Conversely, if the string scale is far from the Planck scale, and daughters are invisible, then this implies that the string completion is strongly coupled.
 
We also check that within our ansatz, these single-trajectory-looking amplitudes do not appear to violate the Veneziano asymptotics of~\cite{Caron-Huot:2016icg}. 

En passant, we complete the Häring-Zhiboedov ansatz by working out the explicit combinatorics of a subset of linear constraints, which allows us to reach larger ansatz sizes.

The main results of our paper are:
\begin{itemize}
\item We prove that weakly coupled gravity amplitudes must have infinitely many trajectories, otherwise they violate the polynomial boundedness of the amplitude.
\item We show concrete evidence of the emergence of a leading sister trajectory at the boundary of coupling space allowed by unitarity of the meromorphic amplitudes (positivity of the residues). We also observe that at the exact limit, on the boundary of coupling space, these solutions become spurious.
\end{itemize}

These results together imply some consequences on the inevitability of strings which complement the existing theorems. 

The paper is organized as follows. In section \ref{sec:setup}, we review the kinematics of the $2\to2$ MHV graviton scattering amplitude and define the assumptions: weak coupling (meromorphy), unitarity (positivity of residues), crossing symmetry, and dispersion relations. Section \ref{sec:theorem} contains the no-go theorem, which is the main analytic result of this work. 
We also explicitly construct a simple example of a single-trajectory amplitude that satisfies all axioms except polynomial boundedness, exhibiting the expected exponential growth. In section \ref{sec:primal}, we present our numerical analysis, based on the Häring-Zhiboedov ansatz, and present our findings regarding the emergence of the ``sister trajectory'' and the divergent behavior of extremal single-trajectory-like solutions in the limit of a large basis size. Finally, section \ref{sec:discussion} provides a discussion of our results and future directions. Technical details regarding Wigner polynomials, an alternative proof of the theorem using partial wave projections, and further visualizations of the numerical data are collected in the appendices \ref{app:Wigner} to \ref{app:state-removal}.



\section{Setup and review}
\label{sec:setup}
In this section, we introduce the relevant scattering amplitude for this work: the ${2\to2 }$ maximally helicity violating (MHV) graviton amplitude, and recall its basic features. We further specialize the discussion to the case of interest, in which it is UV-completed by the exchange of infinitely many weakly-coupled massive higher spin states, and thus corresponds to a spinning version of a dual model amplitude. We review the properties of these models in connection to crossing, unitarity and analyticity, focusing on the case of external particles with spin, following the conventions of \cite{Bern:2021ppb}.

\subsection{Parametrization of the MHV graviton amplitude}

The main object of interest in this paper is the $2 \to 2$ maximally helicity violating amplitude between gravitons, namely 

\begin{equation}
\mathcal{M}(1^+ 2^+ \to 3^- 4^-):=\mathcal{M}_{++--}(s,t,u),
\end{equation}
where $\pm$ indicates the values $\pm2$ of the helicity of each particle.

A convenient way of parametrizing this amplitude is
\begin{equation}
\label{eq:MHVamp}
    \mathcal{M}_{++--}(s,t,u)=\left([12]\langle34\rangle \right)^4 f(s|t,u),
\end{equation}
where $f(s|t,u)$ is a scalar function symmetric under the exchange of $t$ and $u$, but not $s$. The origin of this parametrization is simple: Transformations leaving a momentum $p$ invariant, i.e.~little group transformations, are implemented in spinor helicity language by rescaling the spinors building up $p$ in opposite ways:
\begin{equation}
    p_{\alpha \dot{\alpha}}=|p\rangle_\alpha [p|_{\dot{\alpha}} \longrightarrow \left( t|p\rangle_\alpha \right) \left( t^{-1} [p|_{\dot{\alpha}} \right)=p_{\alpha \dot{\alpha}}.
\end{equation}
By constructing polarization tensors in terms of these momentum spinors, one can check that this implies that polarization tensors corresponding to helicity $h$ particles must rescale as $\epsilon \to t^{-2j}\epsilon$ under little group transformations. Since a generic amplitude is made up of the external particle polarization tensors dotted into a Lorentz-covariant object that has no little group scaling, this means that an amplitude $\mathcal{M}(1^{h_1},2^{h_2},\dots,n^{h_n})$ will scale as 
\begin{equation}
\mathcal{M}(1^{h_1},2^{h_2},\dots,n^{h_n})\longrightarrow \prod_{i}t_i^{-2 h_i}\mathcal{M}(1^{h_1},2^{h_2},\dots,n^{h_n}).
\end{equation}

With the convention that right-handed spinors $|p]_{\dot{\alpha}}$ scale with weight -1 and left-handed spinors with weight 1 and going back to the case of the $2\to2$ graviton MHV amplitude $\mathcal{M}(1^+ 2^+ \to 3^- 4^-)$ we see that the simplest object having the right little-group covariance is indeed $\left([12]\langle34\rangle \right)^4$, and therefore the most general form for the amplitude is this object times a scalar function $f(s,t,u)$.\footnote{Naturally, other combinations of spinors with the right helicity weight are possible, for example $ \left(\langle12\rangle[34]\right)^{-4}$. However, they can always be converted to $\left([12]\langle34\rangle \right)^4$ times a scalar function using spinor identities.} Moreover, since $\left([12]\langle34\rangle \right)^4$ is invariant under the exchange of particles 3 and 4, Bose symmetry implies that $f(s,t,u)$ must also be invariant under $t \leftrightarrow u$, motivating the notation $f(s|t,u)$.  

Following the same logic, we can build all possible MHV amplitudes between four gravitons. For instance, we have that
\begin{equation}
    \mathcal{M}(1^+ 2^- \to 3^- 4^+):=\mathcal{M}_{+--+}(s,t,u)=\left([14]\langle23\rangle \right)^4 f(u|s,t),
\end{equation}
and
\begin{equation}
    \mathcal{M}(1^+ 2^- \to 3^+ 4^-):=\mathcal{M}_{+-+-}(s,t,u)=\left([13]\langle24\rangle \right)^4 f(t|s,u).
\end{equation}

\subsection{Weakly coupled UV-completions}

We will work under the assumption that the graviton interaction is UV-completed by the weakly coupled exchange of massive, higher spin states in such a way that the high energy behavior of the amplitude satisfies the usual gravitational Regge bounds \cite{Chandorkar:2021viw,Haring:2022cyf}, namely that it admits twice-subtracted dispersion relations:

\begin{equation}
    \lim_{|s| \to \infty} \frac{\mathcal{M}_{++--}(s,t,u)}{|s|^2} =0, \quad t<0.
\end{equation}

As is well known, the tree-level exchange of a spin $J$ particle yields a contribution to the amplitude that grows as $s^J$ at large $s$, and therefore these requirements can only be reconciled if we allow for the exchange of infinitely-many massive higher-spin particles of arbitrarily high spin so that cancelations among their contributions can yield a UV behavior compatible with twice-subtracted dispersion relations. This is famously the case for tree-level string theory amplitudes, and is the defining feature of a larger class of amplitudes referred to as dual model amplitudes.

Weak coupling corresponds to the statement that the amplitude should be a meromorphic function of the Mandelstam variables, with infinitely many poles located on the real axis signaling the exchange of on-shell states.
Choosing the parametrization of the momentum spinors in~\eqref{eq:MHVamp} appropriately (see~\cite{Bern:2021ppb} for an explicit parametrization)  one can see that the spinor prefactor satisfies ${\left([12]\langle34\rangle \right)^4 =s^4}$, and thus meromorphy of the full amplitude is equivalent to the meromorphy of $f(s|t,u)$, which should have the aforementioned analytic structure.

The residue at a given pole is constrained by Lorentz symmetry and unitarity. Consider the pole corresponding to the exchange of a particle of mass $m^2$ and spin $J$ in the $s$-channel in a scattering process in which the $i^{\rm th}$ particle has helicity $\lambda_i$. The residue associated with such an interaction must be
\begin{equation}
    {\rm Res}_{s\to m^2} \mathcal{M}(s,t,u)=- c_{m^2,J}^{\lambda_{12},\lambda_{34}} \, d_{\lambda_{12},\lambda_{34}}^J\left(1+\frac{2t}{m^2}\right), 
\end{equation}
with $\lambda_{ij}:\lambda_i - \lambda_j$, and where $d_{\lambda_1-\lambda_2,\lambda_3-\lambda_4}^J$ is a spin $J$ Wigner polynomial signaling the exchange of a spin $J$ particle, generalizing the usual Legendre polynomials to the case in which the external particles have spin \cite{Hebbar:2020ukp}\footnote{For more details on Wigner polynomials, see appendix~\ref{app:Wigner}.}, while $c^{\lambda_{12},\lambda_{34}}_{m^2,J}$ is the square of the coupling constant between the gravitons with their corresponding helicities and the massive spin $J$ particle. Unitarity requires coupling constants to be real in order to have a hermitian Lagrangian, and therefore it implies that $c^{\lambda_{12},\lambda_{34}}_{m^2,J}$ must be non-negative:
\begin{equation}
    c^{\lambda_{12},\lambda_{34}}_{m^2,J} \geq 0.
\end{equation}

In the more general case in which various particles of different spin are exchanged at the same pole, their different contributions can be disentangled using the orthogonality properties of Wigner polynomials, and the residue can be decomposed into a sum of the corresponding polynomials with positive coefficients. 

\subsection{Crossing symmetry for spinning external states}

Crossing symmetry becomes substantially more involved in the presence of spinning external particles than in the more familiar scalar case, where it reduces to the invariance of the amplitude under permutations of the Mandelstam variables. In this subsection, we review how crossing works for spinning particles, with a special emphasis on how to deal with crossing when evaluating the discontinuities of scattering amplitudes between external states with spin. This is textbook material that can be found, for instance, in~\cite{Weinberg:1995mt, Martin:1970hmp, Itzykson:1980rh}.  A more modern and concise review can be found in~\cite{Hebbar:2020ukp,Bellazzini:2016xrt}.

Consider a generic scattering process $in \to out$, in which the incoming states are particles labeled as $\left\{|\vec{p}_i{}^{\sigma_i} \rangle \right\}_{i=1,\dots,n_i}$ and the outgoing ones are labeled as $\left\{|\vec{k}_j{}^{\sigma'_j} \rangle\right\}_{j=1,\dots,n_f}$, where the $\sigma$'s stand for the helicities of the particles, and we omit in the discussion possible additional indices corresponding to other charges as they are not relevant for the case studied in this work. The amplitude for this process has the form
\begin{equation}
    \mathcal{M}\left( \left\{p_i{}^{\sigma_i} \right\}\to \left\{k_j{}^{\sigma_j} \right\}\right)= \mathcal{O}^\ell\left(\left\{p_i{}^{\sigma_i} ,k_j{}^{\sigma_j} \right\};p_1\right) u_\ell^{\sigma_1}(\vec{p}_1),
\end{equation}
where we have purposely singled out the polarization $u_\ell^{\sigma_1}(\vec{p}_1)$ of the incoming particle 1,  $\ell$ is an index corresponding to the representation of the Lorentz group in which the field associated to particle 1 transforms, and $\mathcal{O}^\ell$ is a function of the momenta and helicities of the particles determined dynamically.

Crossing symmetry relates this amplitude to the amplitude $\mathcal{M}\left( \left\{p_i{}^{\sigma_i}  \right\}|_{i \neq 1} \to \left\{\bar{p}_1{}^{\bar{\sigma}_1},k_j{}^{\sigma_j} \right\}\right)$ of a process in which the antiparticle $\bar{1}$ is in the final state with momentum $\bar{p}_1$ and helicity $\bar{\sigma}$ by
\begin{equation}
\mathcal{M}\left( \left\{p_i{}^{\sigma_i}  \right\}|_{i \neq 1} \to \left\{\bar{p}_1{}^{\bar{\sigma}_1},k_j{}^{\sigma_j} \right\}\right)= \pm  \, \mathcal{O}^\ell\left(\left\{p_i{}^{\sigma_i} ,k_j{}^{\sigma_j} \right\};-\bar{p}_1\right) v_\ell^{\bar{\sigma}_1}(\vec{\bar{p}}_1),
\end{equation}
where $\mathcal{O}^\ell$ is the same function as before but evaluated at $p_1 = - \bar{p}_1$, $v_\ell^{\bar{\sigma}_1}(\vec{\bar{p}}_1)$ is the polarization of the outgoing antiparticle $\bar{1}$, and the sign depends on the statistics of the crossed particle and is always positive in the case of bosonic states.

The CPT theorem relates the polarizations of incoming particles and outgoing antiparticles by
\begin{equation}
    u_\ell^\sigma(\vec{p}) = \eta \, v_\ell^{-\sigma}(\vec{p}),
\end{equation}
where $\eta$ is a phase, and therefore we see that the amplitude for the process with an incoming particle with momentum $p_1$ helicity $\sigma$ is equal to the one for the process with an outgoing antiparticle with momentum $-p_1$ and helicity $-\sigma$, up to a phase.

A scenario in which the crossing properties of helicity states play a crucial role is when considering dispersion relations, where we are required to evaluate the discontinuities of scattering amplitudes along different cuts. However, when spinning external states are involved, crossing symmetry implies that for a given helicity configuration, the amplitude evaluated along the left cut -which corresponds to unphysical kinematics for the considered process- is related to amplitudes for processes in different helicity configurations in physical kinematics. Since this will be important for our purposes, let us proceed by reviewing an example that will be useful later to clarify these ideas.

Rather than dealing with a dispersion relation for the full amplitude ${\mathcal{M}_{++--}(s,t,u)}$, we consider an unsubtracted, fixed $t$ dispersion relation for the function $f(s|t,u)$. This is justified because we assumed that $\mathcal{M}_{++--}$ satisfies twice-subtracted dispersion relations, and ${f(s|t,-s-t)=s^{-4}\mathcal{M}_{++--}(s,t,-s-t)}$, and therefore $f(s|t,-s-t)$ decays faster than $s^{-2}$ at large $s$, for $t <0$.

Representing $f(s|t,-s-t)$ as a contour integral using Cauchy's theorem and performing the standard contour deformations, we find the following representation for $f(s|t,-s-t)$, after dropping the vanishing contributions from the contour at infinity:

\begin{equation}
\label{eq:DispRelf}
\begin{split}
    f&(s|,t,-s-t)=\oint \frac{{\rm d}s'}{2 \pi i} \frac{ f(s' |t,-s'-t)}{s'-s}=\frac{8 \pi G_N}{stu}+|\beta_{R^3}|^2\frac{tu}{s}-|\beta_{\phi}|^2 \frac{1}{s}\\
    &+ \frac{1}{\pi}\int_{-\infty}^{-m^2-t} {\rm d}s'\frac{{\rm Disc}_sf(s'|t,-s'-t)}{s'-s}+ \frac{1}{\pi}\int_{m^2}^\infty {\rm d}s'\frac{{\rm Disc}_sf(s'|t,-s'-t)}{s'-s},
\end{split}
\end{equation}
where the massless poles correspond to the universal infrared sector of the most general parity-preserving gravitational theory: the first term is the usual Einstein term, the second corresponds to the correction to the three-point couplings coming from an $R^3$ term in the action, and the third to the coupling to a massless scalar. The integrals run along the branch cuts starting at the first massive state on the $s$-channel for the right branch cut and on the first massive state on the $u$-channel for the left branch cut, and the discontinuities are defined as
\begin{equation}
    {\rm Disc}_x f(x) = \lim_{\epsilon \to 0} \frac{f(x+ i \epsilon) - f(x- i \epsilon) }{2 i}.
\end{equation}

To evaluate the discontinuity along the right branch cut we express $f(s|,t,-s-t)$ in terms of $\mathcal{M}_{++--}(s,t,-s-t)$, as along the physical $s$-channel cut the discontinuity of the $s$-channel amplitude is related by unitarity to the positive spectral density $\rho_J^{++}(s)$ encoding the exchange of spin $J$ states\footnote{This example considers a general $2\to2$ MHV amplitude, not necessarily tree-level, and therefore the spectral densities can contain contributions from multiparticle states and are thus continuous functions above the multiparticle threshold. We will specialize to the weakly-coupled case later.}. Concretely, the integral along the right cut can be expressed as
\begin{equation}
    \int_{m^2}^\infty {\rm d}s'\frac{{\rm Disc}_sf(s'|t,-s'-t)}{s'-s}= \int_{m^2}^\infty {\rm d}s'  \sum_{J=0}^\infty \frac{1+(-1)^J}{2} \frac{\rho^{++}_J(s') \, d_{0,0}^J(1+\frac{2t}{s'})}{s'^4 (s'-s)},
\end{equation}
with $\rho_J^{++}(s) \geq 0$ for $s \geq 0$. Note that it receives contributions from even spins only due to the $t \leftrightarrow u$ invariance of the $s$-channel process.

We would like to use a similar argument for the left cut. Yet, in this case, the integral runs along negative values of $s$ and therefore outside of the $s$-channel physical region ($s \geq 0, \, t \leq 0, \, u \leq 0$), where unitarity dictates the form of the discontinuity.

This is when crossing symmetry comes to our aid. Noting that along the left cut we are in the kinematical region for physical $u$-channel scattering ($u=-s-t \geq 0, \, s,t\leq 0$), we can apply the crossing rules described above to go from the $s$-channel process $1 \, 2 \to 3 \, 4$ to the $u$-channel process $1 \, \bar{4} \to 3  \, \bar{2}$ by swapping particles 2 and 4. Crossing both particles, we obtain
\begin{equation}
    \mathcal{M}(p_1^+,\, p_2^+ \to p_3^- ,\,p_4^-)=\mathcal{M}(p_1^+ ,\, -p_4^+ \to p_3^-, -p_2^-).
\end{equation}
In the notation in which we fix the ordering of the particles, this reads
\begin{equation}
    \mathcal{M}_{++--}(s,t,u)=\mathcal{M}_{+--+}(u,t,s),
\end{equation}
with
\begin{equation}
    \mathcal{M}_{+--+}(u,t,s)=([14]\langle 23 \rangle)^4 \, f(u|t,s).
\end{equation}
Writing the left cut in terms of $M_{+--+}(-s-t,t,s)$ and changing integration variables from $s'$ to $u'=-s'-t$, we have
\begin{equation}
    \int_{-\infty}^{-m^2-t} {\rm d}s'\frac{{\rm Disc}_sf(s'|t,-s'-t)}{s'-s}= \int_{m^2}^\infty {\rm d}u'  \sum_{J=4}^\infty \frac{\rho^{+-}_J(u') \,d_{4,4}^J(1+\frac{2t}{u'})}{(u'+t)^4 (-s-t-u')},
\end{equation}
where $\rho_{J}^{+-}(u)\geq0$ for $u \geq 0$ is the spectral density for the $u$-channel process, and recieves contributions from $J\geq 4$ only by angular momentum conservation.

Putting the contributions from both branch cuts together, we obtain the following dispersive representation for the function $f(s|t,u)$, written entirely in terms of spectral densities and the couplings among the massless states:
\begin{equation}
\begin{split}
    f(s|,t,-s-t)&=\oint \frac{{\rm d}s'}{2 \pi i} \frac{ f(s' |t,-s'-t)}{s'-s}=\frac{8 \pi G_N}{stu}+|\beta_{R^3}|^2\frac{tu}{s}-|\beta_{\phi}|^2 \frac{1}{s}\\
    &-\frac{1}{\pi} \int_{m^2}^\infty {\rm d}x \bigg( \sum_{J=0}^\infty \frac{1+(-1)^J}{2} \frac{\rho^{++}_J(x) \, d_{0,0}^J(1+\frac{2t}{x})}{x^4 (s-x)} \\ & \hspace{3cm}+\sum_{J=4}^\infty \frac{\rho^{+-}_J(x) \,d_{4,4}^J(1+\frac{2t}{x})}{(x+t)^4 (-s-t-x)} \bigg).
\end{split}
\end{equation}

\section{The theorem}

\label{sec:theorem}

In this section, we prove that weakly-coupled graviton scattering amplitudes consistent with crossing symmetry and unitarity cannot be UV-completed by a single Regge trajectory if we require in addition that the amplitude should be polynomially bounded. 

Polynomial boundedness is the statement that for any finite $t$, there should exist an integer $N(t)$ such that for large $s$, the fixed $t$ behavior of the amplitude satisfies the bound
\begin{equation}
    \lim_{|s|\to \infty} \frac{\mathcal{M}(s,t)}{s^{N(t)}} \to 0.
\end{equation}

It is a famous fact that an amplitude that grows exponentially in (some direction of) the complex $s$ plane leads to violations of causality of the crudest form: it allows for backwards in time propagation in position space (see~\cite{iagolnitzer:hal-00155691} for the proof in axiomatic QFT, or~\cite{Camanho:2014apa} for a modern review of the argument in a toy model in $0+1$-dimensions.). Thus, causality requires that scattering amplitudes must be bounded by any exponential, namely
\begin{equation}
    \lim_{|s|\to \infty} \frac{\mathcal{M}(s,t)}{e^{k |s|}} \to 0, \quad \forall k \geq 0.
\end{equation}

Polynomial boundedness is a stronger condition that is expected on very general grounds when, on top of causality, we ask for the standard features of a healthy quantum theory: unitarity and local interactions~\cite{iagolnitzer:hal-00155691,Eden:1971fm,Epstein:1969bg,Haring:2022cyf}. It is a key property underlying many of the most important results in $S$-matrix theory, such as the Froissart-Martin bound on the growth of the cross section~\cite{Froissart:1961,Martin:1963}. Interestingly, in non-local field theories, this property is relaxed, and generically one finds amplitudes that are sub-exponential yet not polynomially bounded~\cite{Efimov:1969zjc,Alebastrov:1973np}. This is the case, e.g., in so-called BKGM gravity, a non-local generalization of GR with infinite-derivative interactions~\cite{Biswas:2011ar,Biswas:2013kla,Talaganis:2014ida}.

In what follows, we will show that starting from the assumptions outlined in the previous section (tree-level interactions, twice-subtracted fixed $t$ dispersion relations, unitarity and crossing symmetry), a $2\to 2$ graviton amplitude UV-completed by a spectrum composed of a single Regge trajectory grows faster than any polynomial at large $s$. The heart of the proof is that crossing symmetry fixes the asymptotic decay of the couplings between the graviton and the exchanged massive higher spin particles at large mass levels. This behavior, when injected back into the amplitude, yields non-polynomially bounded growth.

We focus on a single trajectory, as the generalization to the case of any finite number of trajectories is trivial. We comment on this at the end of the proof.

\subsection{Asymptotic decay of the couplings from crossing symmetry}

The first step is to obtain the asymptotic behavior of the coupling between the external graviton and the massive higher-spin particles composing the trajectory at large energies.

To do this, we start from the dispersive representation of the $f(s|t,u)$ function derived in the previous section:

\begin{equation}
\label{eq:DispRelf}
\begin{split}
    f(s|,t,-s-t)&=\oint \frac{{\rm d}s'}{2 \pi i} \frac{ f(s' |t,-s'-t)}{s'-s}=\frac{8 \pi G_N}{stu}+|\beta_{R^3}|^2\frac{tu}{s}-|\beta_{\phi}|^2 \frac{1}{s}\\
    &-\frac{1}{\pi} \int_{m^2}^\infty {\rm d}x \bigg( \sum_{J=0}^\infty \frac{1+(-1)^J}{2} \frac{\rho^{++}_J(x) \, d_{0,0}^J(1+\frac{2t}{x})}{x^4 (s-x)} \\ & \hspace{3cm}+\sum_{J=4}^\infty \frac{\rho^{+-}_J(x) \,d_{4,4}^J(1+\frac{2t}{x})}{(x+t)^4 (-s-t-x)} \bigg),
\end{split}
\end{equation}

This expression can be specialized to the case of a weakly-coupled amplitude by noting that if we allow only for tree-level interactions, then the discontinuity of the amplitude will be made of a collection of delta functions at the locations of the poles in the amplitude. Namely, the spectral densities $\rho^{++}_J(x)$ and $\rho^{+-}_J(x)$ will be of the form
\begin{equation}
    \rho^{++}_{J}(x)=- \pi \sum_{n} c^{++}_{n,J} \, \delta\left(x-m_{n,J}^2\right), \quad  \rho^{+-}_{J}(x)=- \pi \sum_n c^{+-}_{n,J} \, \delta\left(x-m_{n,J}^2\right),
\end{equation}
where for each value of $J$ the sum in $n$ runs over all the particles with spin $J$ in the spectrum.

Now, we restrict the spectrum to be composed of a single Regge trajectory. This means that there should be only one particle in the spectrum with a given value of $J$. Thus, in the case of a single trajectory composed of even spins, the spectral densities are simply
\begin{equation}
    \rho^{++}_{2n}(x)=- \pi c^{++}_{n,2n+2} \, \delta\left(x-m_n^2\right), \quad  \rho^{+-}_{2n}(x)=- \pi c^{+-}_{n,2n+2} \, \delta\left(x-m_n^2\right),
\end{equation}
and the dispersive representation of eq.~\eqref{eq:DispRelf} reduces to
\begin{equation}
\label{eq:DispRelsingletraj}
    \begin{split}
    f(s|,t,-s-t)&=\oint \frac{{\rm d}s'}{2 \pi i} \frac{ f(s' |t,-s'-t)}{s'-s}=\frac{8 \pi G_N}{stu}+|\beta_{R^3}|^2\frac{tu}{s}-|\beta_{\phi}|^2 \frac{1}{s}\\
    &+ \sum_{n=1}^\infty \bigg( \frac{c^{++}_{n,2n+2} \, d_{0,0}^{2n+2}(1+\frac{2t}{m_n^2})}{m_n^8 (s-m_n^2)} + \frac{c^{+-}_{n,2n+2}\,d_{4,4}^{2n+2}(1+\frac{2t}{m_n^2})}{(m_n^2+t)^4 (-s-t-m_n^2)} \bigg).
\end{split}
\end{equation}

The key observation is that, while the function $f(s|t,u)$ is symmetric under the exchange of $t$ and $u$, eq.~\eqref{eq:DispRelsingletraj} does not exhibit this symmetry manifestly. Rather, $t \leftrightarrow u$ symmetry is a constraint on the masses and couplings of a consistent amplitude. 

In particular, eq.~\eqref{eq:DispRelsingletraj} exhibits infinitely many poles in $u$, but the only explicit pole in $t$ is the pole at $t=0$. Yet, we expect that the dispersion relation used to obtain~\eqref{eq:DispRelsingletraj} should be valid up to $t=m_1^2$, the mass of the lightest massive state, and therefore the pole at $t=m_1^2$ must be generated by the divergence of the infinite sum in~\eqref{eq:DispRelsingletraj}. 

We can argue that~\eqref{eq:DispRelsingletraj} is valid up to $t=m_1^2$, by noting that at large $s'$ the integrand behaves as
\begin{equation}
   \int {\rm d}s' \frac{f(s'|t,u)}{s'-s} \sim \int {\rm d}s' \frac{s'^{-4}M_{++--}(s',t)}{s'} \sim \int {\rm d}s' s'^{\alpha(t)-5},
\end{equation}
and since $\alpha(t=m_1^2)=4$ by definition, we see that as $t \to m_1^2$ the integrand behaves as $s^{-1}$, yielding a divergent integral, while for $t<m_1^2$ the integrand is bounded by $s^{-1}$ and therefore the integral converges.\footnote{This assumes that the trajectory function $\alpha(t)$ is monotonic, which seems a very mild assumption. Yet, even if $\alpha(t)$ was not monotonic and there was a value $t_* < m_1^2$ such that $\alpha(t_*)=4$, then the dispersion relation in eq.~\eqref{eq:DispRelsingletraj} would hold up to $t=t_*$. Replacing $m_1^2$ by $t_*$ in what follows leads to the same conclusion, but we choose to work with $m_1^2$ as the limit of the region of validity of the dispersion relation, as it appears more natural.} 

To investigate whether the sum in~\eqref{eq:DispRelsingletraj} can produce a pole at $t=m_1^2$ we must study the large $n$ asymptotics of the series $\mathcal{S}(s,t)$,
\begin{equation}
    \mathcal{S}(s,t):=\sum_{n=1}^\infty \bigg( \frac{c^{++}_{n,2n+2} \, d_{0,0}^J(1+\frac{2t}{m_n^2})}{m_n^8 (s-m_n^2)} + \frac{c^{+-}_{n,2n+2}\,d_{4,4}^J(1+\frac{2t}{m_n^2})}{(m_n^2+t)^4 (-s-t-m_n^2)} \bigg).
\end{equation}
At large $J$, Wigner polynomials behave as
\begin{equation}
   d_{\lambda\lambda'}^J(z)\sim(-1)^{\lambda'-\lambda}\sqrt{\frac{1}{\pi J }}\left(\frac{2}{z-1}\right)^{1/4}\cos\left(iJ \sqrt{2(z-1)}-\frac{J \pi}{2} \right) \sim e^{J\sqrt{2(z-1)}},
\end{equation}
and therefore we have that at large $n$, $d_{\lambda\lambda'}^{2n+2} (1+\tfrac{2t}{m_n^2})$ behaves as
\begin{equation}
        d_{\lambda\lambda'}^{2n+2} \left(1+\frac{2t}{m_n^2}\right)= \, e^{4n \sqrt{t/m_n^2} } \quad \text{if} \, m_n^2 < n^2 \, {\rm as} \, \, n \to \infty.
\end{equation}
When $m_n^2$ is not bounded by $n^2$ at large $n$ it can be seen that $d_{\lambda\lambda'}^{2n+2} (1+\tfrac{2t}{m_n^2})$ tends to 1 at large $n$, independently of the value of $t$. Since in that case it is impossible to satisfy crossing, spectra not bounded by $n^2$ are ruled out from the start. Focusing from now on on spectra such that $m_n^2 < n^2$ asymptotically, we have that the large $n$ tail of the series behaves as
\begin{equation}
    \mathcal{S}(s,t)\sim \sum_n \frac{c^{++}_{n,2n+2}+c^{+-}_{n,2n+2}}{-m_n^{10}}e^{4n \sqrt{t/m_n^2}}.
\end{equation}

Since all the $c^{++}_{n,2n+2}$ and $c^{+-}_{n,2n+2}$ couplings are non-negative by unitarity, demanding that the sum is finite for $t<m_1^2$ and diverges precisely at $t=m_1^2$ yields the following asymptotic decay for the sum of the couplings:
\begin{equation}
    \label{eq:asymptoticdecay}
    c^{++}_{n,2n+2}+c^{+-}_{n,2n+2} \sim e^{-4n \sqrt{m_1^2/m_n^2}}.
\end{equation}
which implies that at least one of the two families of couplings must decay asymptotically at this speed, while the other could decay at this speed or faster but crucially not slower, as this would cause the amplitude to diverge for values of $t$ below $m_1^2$.

In what follows, we will see that this decay yields an amplitude whose high-energy behavior always violates polynomial boundedness.

\subsection{Non-polynomial high-energy behavior of single-trajectory amplitudes}

Now we proceed to showing that the decay of the coefficients obtained by imposing crossing yields an amplitude whose high-energy behavior is not bounded by any polynomial.

A simple way of testing whether an amplitude (or its discontinuity) is bounded at large $s$ by $\mathcal{M}(s,t,-s-t)< s^{N(t)}$ as $s\to \infty$ for some $N(t)$ is to consider the following integral:
\begin{equation}
\label{eq:testintegral}
  \mathcal{I}^{(N)}(t) = \int_{m^2}^\infty {\rm d}s \frac{{\rm Disc_s}\mathcal{M}(s,t)}{s^{\bar{N}+1}},
\end{equation}
where $\bar{N}$ is the smallest integer larger that $N(t)$, with $t$ fixed, and where $m^2$ is some suitably defined energy scale. Naturally, if the amplitude is bounded by $s^{N(t)}$, then $\mathcal{I}^{(N)}(t)$ will be finite. On the other hand, if the amplitude grows faster than $s^{N(t)}$ at large $s$, fixed $t$, then the integral in~\eqref{eq:testintegral} will diverge.

Let us assume that we choose an $N(t)$ such that the integral in eq.~\eqref{eq:testintegral} converges. If this is the case, then we can safely exchange the integral in~\eqref{eq:testintegral} and the sum over states contained in the discontinuity of the amplitude. In particular, specializing~\eqref{eq:testintegral} to the case of tree-level MHV scattering of gravitons with a single Regge trajectory and taking as $m^2$ the mass $m_1^2$ of the first massive state, the requirement that the amplitude be bounded by $s^{N(t)}$ is translated to the convergence of the following series:
\begin{equation}
\label{eq:testseries}
 \mathcal{I}^{(N)} (t)=   \sum_{n=1}^\infty  \frac{c^{++}_{n,2n+2} \, d_{0,0}^{2n+2}(1+\frac{2t}{m_n^2})}{(m_n^2)^{\bar{N}+1} }.
\end{equation}

Yet, exactly as when requiring the presence of the crossed channel pole in the previous subsection, we can analyze the convergence of~\eqref{eq:testseries} by using the large $n$ behavior of the Wigner polynomials as well as the requirement that at least one of the two families of couplings decays as $\exp(-4n\sqrt{m_1^2/m_n^2})$ (recall that from the condition~\eqref{eq:asymptoticdecay} it is enough that either the $\{ c^{++}_{2n+2}\}$ or the $\{c^{+-}_{2n+2}\} $ decay this way, but not necessarily both - the other set of couplings could decay faster).

Let us assume first that it is the $c^{++}_{2n+2}$ set of coefficients that decays this way. In that case, the asymptotic behavior of the series is
\begin{equation}
\label{eq:polybound-series}
     \mathcal{I}^{(N)} (t) \sim  \sum_{n}^\infty  \frac{e^{4n (\sqrt{t/m_n^2}-\sqrt{m_1^2/m_n^2})}}{(m_n^2)^{\bar{N}} },
\end{equation}
which is divergent for {\bf any} $\bar{N}$ if $t \geq m_1^2$, and therefore we conclude that the large $s$, fixed $t$ behavior of the amplitude is not bounded by any polynomial in $s$. 

If, on the other hand, it were the $\{c^{+-}_{2n+2}\} $ satisfying $\{c^{+-}_{2n+2}\} \sim \exp(-4n\sqrt{m_1^2/m_n^2})$, we could repeat the same argument but using the discontinuity in the $u$-channel to probe the UV behavior of the amplitude\footnote{Or equivalently the $s$-channel discontinuity integrated along the negative real axis.}, in this way we obtain a condition analogous to eq.~\eqref{eq:polybound-series} but involving the $u$-channel coefficients $c^{+-}_{2n+2}$ when integrating it along the positive, real $u$-axis, which can be seen to be divergent for any choice of $N(t)$ in the same way, concluding the proof.

\subsection{Extension to any finite number of trajectories}

Extending the previous result from the case with a single Regge trajectory to a scenario with any finite number of them is straightforward. The generalization of the single-trajectory dispersive representation for $f(s|t,u)$ in eq.~\eqref{eq:DispRelsingletraj} to the case with $K+1$ Regge trajectories is simply
\begin{equation}
\label{eq:DispRelmanytraj}
    \begin{split}
    &f(s|,t,-s-t)=\frac{8 \pi G_N}{stu}+|\beta_{R^3}|^2\frac{tu}{s}-|\beta_{\phi}|^2 \frac{1}{s}\\
    &+ \sum_{i=0}^K\sum_{n=1}^\infty \bigg( \frac{c^{++}_{n,2n+2-2i} \, d_{0,0}^{2n+2-2i}(1+\frac{2t}{m_n^2})}{m_n^8 (s-m_n^2)} + \frac{c^{+-}_{n,2n+2-2i}\,d_{4,4}^{2n+2-2i}(1+\frac{2t}{m_n^2})}{(m_n^2+t)^4 (-s-t-m_n^2)} \bigg),
\end{split}
\end{equation}
with $c^{++}_{n,2n-2-2i}=c^{+-}_{n,2n-2-2i}=0$ if $2n-2-2i<0$, namely the spectrum at level $n$ comprises particles of spins $2n+2, 2n, 2n-2,\dots2n-2K$.

In order for~\eqref{eq:DispRelmanytraj} to reproduce the pole at $t=m_1^2$ while staying finite, the couplings between the graviton and the massive states must satisfy is
\begin{equation}
    c^{++}_{n,2n+2-2i}+c^{+-}_{n,2n+2-2i} \leq e^{-4n \sqrt{m_1^2/m_n^2}}, \quad \forall i=0,\dots,K, 
\end{equation}
and
\begin{equation}
\label{eq:asymptdecay-manytraj}
    c^{++}_{n,2n+2-2i}+c^{+-}_{n,2n+2-2i} \sim e^{-4n \sqrt{m_1^2/m_n^2}}, \quad \text{for at least one }i \in 0,\dots,K.
\end{equation}
The essential distinction from the single--Regge-trajectory case is that we no longer require a strict decay condition such as~\eqref{eq:asymptoticdecay} to hold for \emph{every} trajectory. Instead, an analogous decay must be satisfied by \emph{at least one} of them in order to generate the pole at $t = m_1^2$. This follows from the observation that, as $t \to m_1^2$, the sum of the contributions from all trajectories must diverge like $(t - m_1^2)^{-1}$. Because the number of trajectories is finite, such a divergence can only occur if at least one individual contribution diverges, but it is not necessary for all of them to do so. 

Once we have the decay condition~\eqref{eq:asymptdecay-manytraj} imposed by crossing in the case of many trajectories, the proof follows exactly as in the single trajectory case: When integrating the discontinuity of amplitude the amplitude times $s^{-N(t)}$ as in eq.~\eqref{eq:testintegral}, the contribution from the trajectory that saturates the decay condition~\eqref{eq:asymptdecay-manytraj} will diverge for any $N(t)$ on the same grounds as in the previous case, showing that any finite number of trajectories cannot cure the problem.

Finally, let us comment on the case of infinitely many trajectories. Naturally, we know many examples of tree-level UV-completions of graviton scattering amplitudes involving infinitely many trajectories, so this case must evade the no-go presented above. The reason why it does so is simple: having infinitely many trajectories, the condition~\eqref{eq:asymptdecay-manytraj} is not required anymore, as even if the contribution from each trajectory is finite as $t\to m_1^2$, the sum over infinitely many trajectories can diverge and provide the expected singularity.

\subsection{Example of exponentially growing single-trajectory amplitude}

In the previous sections, we showed that an amplitude consisting of a single Regge trajectory cannot be compatible with crossing and unitarity if it is required to be polynomially bounded -- and therefore causal -- at large energies. 

On the other hand, if we relax causality, it is easy to construct crossing symmetric, unitary amplitudes having a single trajectory as their spectrum. For simplicity, rather than working with gravitons, consider the scattering of color-ordered scalars, such that the amplitude is only symmetric under the exchange of $s$ and $t$, and possesses only $s$ and $t$-channel poles, but no $u$-channel poles.

Assuming moreover a linear spectrum, a single-trajectory amplitude of this type would have infinitely many poles -one at each non-negative integer value of $s$ and $t$, with residues given by
\begin{equation}
    {\rm Res}_{s\to n}\mathcal{A}(s,t)= - c_n \mathcal{P}_n \left(1+\frac{2t}{n}\right), \quad {\rm Res}_{t\to n}\mathcal{A}(s,t)= - c_n \mathcal{P}_n \left(1+\frac{2s}{n}\right), \quad c_n \geq 0,
\end{equation}
where the $\mathcal{P}_n$ are Legendre polynomials.

Now, recall that the constraint on the large $n$ behavior of the couplings came from the requirement that the $s$-channel pole expansion should reproduce the leading $t$-channel pole, which fixed the decay of the $c_n$. However, the reason why we had to impose this constraint was that having a polynomial bounded amplitude implied the existence of dispersion relations, which prevented both $s$ and $t$-channel poles from appearing explicitly in the expression for the amplitude. Yet, if we abandon the requirement that the amplitude should not grow exponentially at large energies, then any amplitude of the form
\begin{equation}
\label{eq:single-traj-exp}
    \mathcal{A}(s,t)= -\sum_n c_n \left( \frac{\mathcal{P}_n(1+\tfrac{2t}{n})}{s-n}+\frac{\mathcal{P}_n(1+\tfrac{2s}{n})}{t-n}\right), \quad c_n \geq 0,
\end{equation}
gives a good single-trajectory amplitude, as long as the $c_n$ are chosen such that~\eqref{eq:single-traj-exp} converges for any finite $s,t$. Since the large $n$ behavior of the Legendre polynomials is also given by eq.~\eqref{eq:Wignerdasymptotic} ($\mathcal{P}_n$ corresponds to a Wigner polynomial with all helicities set to 0, i.e. $\mathcal{P}_n(z)=d_{00}^n(z)$), for a linear spectrum the sum converges if
\begin{equation}
    \lim_{n \to \infty} c_n \, e^{2\sqrt{n}t } =0, \quad \forall \,t,
\end{equation}
which is solved explicitly by taking for example $c_n= e^{-n}$. Therefore, 
\begin{equation}
 \label{eq:single-traj-sol}
    \mathcal{A}(s,t)= -\sum_n e^{-n} \left( \frac{\mathcal{P}_n(1+\tfrac{2t}{n})}{s-n}+\frac{\mathcal{P}_n(1+\tfrac{2s}{n})}{t-n}\right),
\end{equation}
is crossing-symmetric, unitary and finite for any finite value of $s$ and $t$ away from the poles, and therefore corresponds to a valid single-trajectory amplitude if one does not enforce any bounds on its high-energy behavior. 

It is simple to see that this amplitude violates causality. By a direct evaluation of the amplitude at large $s$, fixed $t$, we can check that it grows exponentially with $s$. 

When studying the large $s$ behavior of~\eqref{eq:single-traj-sol}, we can focus on the contribution from the $t$-channel poles, as the one coming from the $s$-channel poles simply decays as $s^{-1}$. We can estimate the high-energy behavior of the $t$-channel contribution analytically by approximating the sum by an integral and replacing the Legendre polynomials by their asymptotic behavior at large $n$ and argument $z:=\cosh\eta$ larger than 1:
\begin{equation}
    \mathcal{P}_n(\cosh\eta)\sim \frac{e^{(n+\frac{1}{2})\eta}}{\sqrt{2 \pi n \sinh \eta}}, \quad {\rm for}\cosh\eta=z>1,
\end{equation}
which is basically a particular case of eq.~\eqref{eq:Wignerdasymptotic}.

We can thus approximate
\begin{equation}
    -\sum_n \frac{e^{-n}\mathcal{P}_n(1+\tfrac{2s}{n})}{t-n} \sim \int {\rm d}n  \frac{e^{-n+(n+\frac{1}{2})\eta(s,n)}}{\sqrt{2 \pi n \sinh \eta(s,n)}}, \quad \cosh \eta = 1+\tfrac{2s}{n}
\end{equation}
and provide an estimate of the large $s$ behavior of the amplitude by approximating the integral by a saddle. 

\begin{figure}
    \centering
    \includegraphics[width=0.5\linewidth]{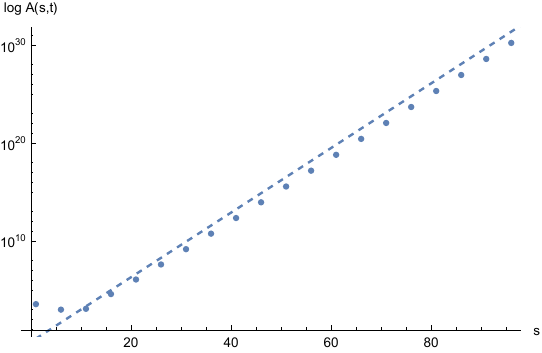}
    \caption{Blue dots: Numerical evaluation of~\eqref{eq:single-traj-sol} for $t=0$. Saddle point approximation, $e^{0.76s}$.}
    \label{fig:singletrajexp}
\end{figure}

At large $s$, the location of the saddle occurs for values of $n$ of order $s$, and therefore it is sensible to rescale $n=\alpha s$, such that $\eta$ depends only on $\alpha$, but not on $s$ and $n$ individually. Changing integration variables to $\alpha$ and solving numerically for the position $\alpha_*$ of the saddle in $\alpha$  we find (setting $t=0$ for simplicity):
\begin{equation}
    \mathcal{A}(s,0) \sim - \int {\rm d}\alpha  \,s \frac{e^{-\alpha s+(\alpha s+\frac{1}{2})\eta(\alpha)}}{\sqrt{2 \pi \alpha s \sinh \eta(\alpha)}} \sim e^{s\alpha_*(\eta_*-1)}\sim e^{0.76s},
\end{equation}
with $\alpha_* \sim 1.11$ and  $\eta_*:=\eta(\alpha_*)\sim1.69$, exhibiting clear exponential behavior that matches the numerical evaluation of the amplitude very well (see fig.~\ref{fig:singletrajexp}).

In any case, the relevant message is that when giving up polynomial boundedness, it is a simple matter to build amplitudes with a single Regge trajectory. The price to pay is that these amplitudes grow exponentially with energy, and therefore violate causality.

\section{Primal bootstrap}
\label{sec:primal}

In this section, we present the results of a numerical analysis addressing the question that was posed and examined in \cite{Albert:2024yap}.\footnote{See also the related deformation to the Veneziano amplitude proposed in~\cite{Arkani-Hamed:2020blm}.} 
Following the ansatz introduced in \cite{Haring:2023zwu} for graviton scattering, generalizing an old open-string construction by Khuri~\cite{Khuri:1967cmn,Khuri:1969elr}, we show how maximizing couplings results in explicit amplitudes that exhibit single-trajectory-like behavior at low masses. We also show that the limit for the extremal amplitudes where the ansatz size goes to infinity is not well defined, as the parameters of the ansatz diverge and yield an infinite result for the amplitude. Yet, the finite ansatz bounds are still meaningful as their convergence with the size of the ansatz is very good, and thus the bounds themselves have a well-defined limit even if the extremal amplitudes do not.

En passant, we complete the ansatz of \cite{Haring:2023zwu} by explicitly solving a set of linear constraints, reducing the number of parameters. This simplifies the use of the ansatz for numerics at a large number of parameters.

\subsection{Ansatz definition}
\label{sec:numerics-ansatz}

To set the stage, we briefly review the setting and ansatz presented in \cite{Haring:2023zwu} for tree-level graviton scattering in the case of integer spectra, on which our numerical exploration is based. This anatz parametrizes the $2\rightarrow2$ MHV scattering amplitude in Eq.~\ref{eq:MHVamp} as follows:
\begin{align}
\label{eq:primal-ansatz}
    f(s|t,u) = -8\pi G_N &\frac{\Gamma(-s)\Gamma(-t)\Gamma(-u)}{\Gamma(1+s)\Gamma(1+t)\Gamma(1+u)}\nonumber\\
    &+ \sum_{(c_s,c_{tu}, d_s, d_{tu})\in \mathcal{I}}\alpha_{(c_s,c_{tu}, d_s, d_{tu})}\frac{\Gamma(c_s-s)\Gamma(c_{tu}-t)\Gamma(c_{tu}-u)}{\Gamma(d_s+s)\Gamma(d_{tu}+t)\Gamma(d_{tu}+u)}\rm{.}
\end{align}

As derived in \cite{Haring:2023zwu}, the tuple $(c_s,c_{tu}, d_s, d_{tu})$ contained in the set of all admissible combinations $\mathcal{I}$ fulfills several consistency relations to ensure polynomial residues and the correct behavior of the leading trajectory ($j(t) = 2t +2$) in the Regge limit. In addition, there remain additional linear dependencies among the admissible terms in $\mathcal{I}$ that we need to remove before working with the ansatz, mixing the gamma functions around, coming from the Gamma identity $\Gamma(z+1) = z \Gamma(z)$ and using $s+t+u=0$. In~\cite{Haring:2023zwu}, it was sufficient to numerically solve 
the dependencies for each level successively in $N_{\mathrm{max}}$ for incrementing $c_s = 0,\ldots,N_{\mathrm{max}}$. This renders   $N_{\rm max}=20-21$ accessible.\footnote{We thank Kelian H\"aring for a discussion on this and sharing a mathematica notebook with some code and solutions to the linear dependencies.}

Attempting to reach higher values of $N_{\rm max}$ in our analysis, we found that this problem can be solved analytically in a closed form, which we briefly explain now. The linear dependencies occur among three distinct subsets of parameter tuples:
\begin{enumerate}
    \item $\{(c_s,c_{tu}, d_s,d_{tu}), (c_s-1,c_{tu}, d_s,d_{tu}),(c_s-1,c_{tu}, d_s-1,d_{tu})\}\subseteq\mathcal{I}$,
    \item $\{(c_s,c_{tu}, d_s,d_{tu}), (c_s,c_{tu}-1, d_s-1,d_{tu}),(c_s,c_{tu}-1, d_s,d_{tu}-1), (c_s,c_{tu}-1, d_s,d_{tu}) \}\subseteq\mathcal{I}$ with $c_s+d_s\geq3$,
    \item $\{(c_s,c_{tu}, d_s,d_{tu}), (c_s,c_{tu}-1, d_s,d_{tu}-1),(c_s+1,c_{tu}-1, d_s,d_{tu}) \}\subseteq\mathcal{I}$ with $c_s+d_s\neq2$.
\end{enumerate}
The validity of the linear dependence among the elements of these subsets can be checked in a straightforward computation. In fact, the first subset is solely the consequence of $\Gamma(z+1) = z\Gamma(z)$, while the other two subsets arise in the specific setting of graviton scattering as a consequence of momentum conservation in the form of the relation~${s+t+u=0}$ between the Mandelstam variables on top of the gamma function identities. In practice, to generate a set of fully linearly independent ansatz terms, we scan $\mathcal{I}$ and remove redundant elements sequentially: we first resolve the dependencies of type 1, followed by type 2, and finally type 3. We conclude that this is a sufficient condition to remove all dependencies because eventually $|\mathcal{I}|=N_{\mathrm{max}}^2+3N_{\mathrm{max}}-2$, which is precisely the relation found in \cite{Haring:2023zwu} when removing the linear dependencies numerically. 

By construction, this ansatz fulfils most of the properties of a sane scattering amplitude except for unitarity. In what follows, we numerically enforce unitarity of the amplitude in full analogy to the approach of \cite{Haring:2023zwu}. This requires us to impose positivity of the coefficients $c_{n,J}^{++}$ and $c_{n,J}^{+-}$ at each residue in the two scattering channels $(++\rightarrow++)$ and $(+-\rightarrow +-)$ (see, e.g., the single trajectory definition in Eq.~\ref{eq:DispRelsingletraj}). In this setup, we will implement a primal bootstrap approach to study the properties of scattering amplitudes arising from maximizing one of the occurring couplings, depending on the number of ansatz terms controlled by $N_{\mathrm{max}}$. To this end, we utilize the numerical optimization routines offered by the software package SDPB \cite{Simmons-Duffin:2015qma, Landry:2019qug}, for which we created a convenience wrapper~\cite{Metayer:SDPBwrapper} to facilitate its usage in \texttt{mathematica} and its distribution on computational resources. 

As a final comment, note that a nice feature of this ansatz is that once unitarity is imposed in a sufficient number of residues for some value of $N_{\rm max}$, one can check that it is also satisfied in all remaining residues. Thus, the obtained extremal amplitudes at finite $N_{\rm max}$ are fully consistent tree-level amplitudes satisfying all the required axioms.

\subsection{Numerical results}
\label{sec:numerics-results}

Let us now present the results of this primal bootstrap analysis.

\subsubsection{The sister}

Most of the results of the numerical analysis refer to the case where we maximize the coupling $c_{1,4}^{++}$ sitting on the leading trajectory (the amplitude is evaluated in units of $8\pi G_N$). In the left panel of Fig.~\ref{fig:amplitude-main}, we show the obtained amplitude for $N_{\rm max}=20$, which only lists couplings $c_{n,J}^{++}$ with even spins.\footnote{A more complete series of such plots for increasing $N_{\rm max}$ is provided in fig.~\ref{fig:couplings-gif} in appendix \ref{app:ampl-gig}.}
We can identify the emergence of a \textit{sister} trajectory with slope one-half of the leading trajectory, $j_{\rm sister}(t)\sim t$ that towers over the other couplings. Strikingly, for masses up to order $n\sim20$ the couplings between the graviton and states below the sister trajectory are completely switched off, while the couplings to the states above it are non-zero but orders of magnitude smaller with respect to the ones of the emergent sister trajectory. 

This hierarchy between the couplings in the sister trajectory and the rest causes the amplitude to effectively behave as a single-trajectory object, at least at energies below the mass at which the states below the trajectory are switched on. What is more, we checked that this sister is indeed the dominating trajectory of the amplitude in the Regge limit, as we show in sec.~\ref{sec:numerics-regge-asymptotics}. 

\begin{figure}
    \centering
    \includegraphics[width=\textwidth]{amplitudeTable_Nmax_20_MAX_c14_80_Poles_with_HorizontalSlice.pdf}
    \caption{We show in the left panel the first couplings $c_{n,J}^{++}$ obtained by maximizing the value of $c_{1,4}^{++}$ at $N_{\rm max} = 20$. We clearly identify an emerging sister trajectory that dominates the couplings by many orders of magnitude (note the log-scale of the couplings). The right panel visualizes a horizontal slice through the coupling matrix at the level $n=15$ at $N_{\rm max} = 20$ (green, see also the box in the left panel) and the corresponding values of the Virasoro-Shapiro amplitude's couplings (orange). Note that the upper plot is in log-scale, whereas the lower plot uses a linear scale for the couplings' values. We observe the growth of the couplings related to the emergent sister trajectory around $J\sim 14$.}
    \label{fig:amplitude-main}
\end{figure}

\begin{figure}
    \centering
    \includegraphics[width=\textwidth]{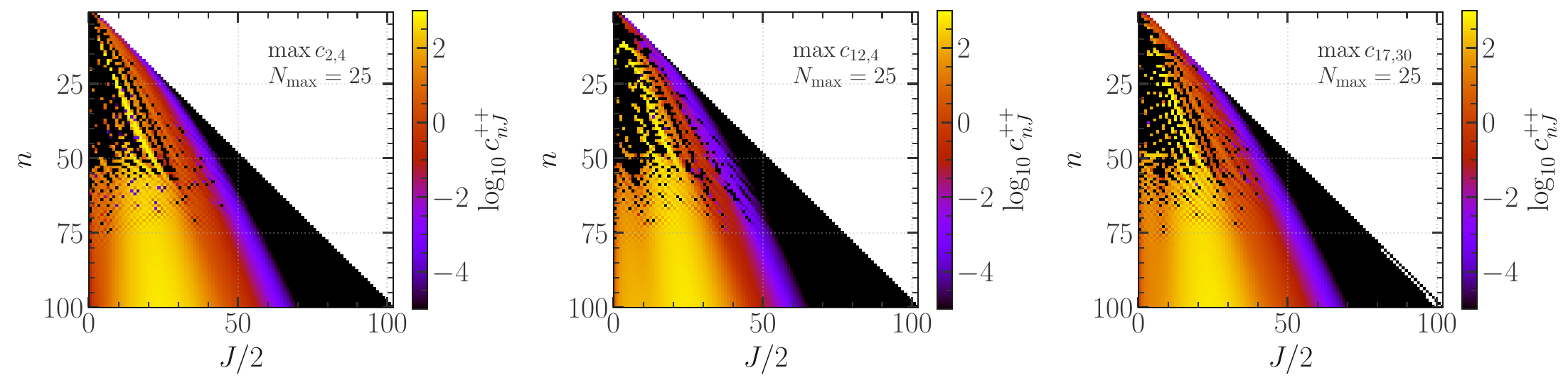}
    \caption{We show the emergence of the sister trajectory with unit slope also when maximizing the coefficients $c^{++}_{2,4}$, $c^{++}_{12,4}$ and $c^{++}_{17,30}$ as an indication of the robustness of the phenomenon.}
    \label{fig:max-alt-c}
\end{figure}

In fig.~\ref{fig:max-alt-c} we show that the emergence of a sister trajectory seems to be a robust phenomenon regardless of which coupling one chooses to maximize by presenting plots analogous to the one in fig.~\ref{fig:amplitude-main} but maximizing the coefficients $c^{++}_{2,4}$, $c^{++}_{12,4}$ and $c^{++}_{17,30}$ rather than $c^{++}_{1,4}$. 

Lastly, we also illustrate the space that is maximally allowed by this ansatz via the example of $c_{1,4}^{++}$ and $c_{2,0}^{++}$ in appendix \ref{app:almond}.

\subsubsection{Increasing the size of the ansatz}

In this subsection we study the behavior of the extremal solutions as we increase the size $N_{\rm max}$ of the ansatz.

In fig.~\ref{fig:couplings-gif} we show the values of the couplings of the amplitude resulting from the maximization of $c_{1,4}^{++}$ for values of $N_{\rm max}$ from 3 to 30.

The sequence of plots reveals a striking sharpening of the spectral features as the number of parameters increases. At low $N_{\rm max}$, the couplings are distributed somewhat broadly. However, as $N_{\rm max}$ becomes large, the solution develops a specific structure characterized by the emergence of the mentioned sister trajectory. 

Crucially, the plots demonstrate a progressive suppression of the couplings for states located below this sister trajectory (the lower-left region of the spectral triangle) up to a given mass threshold that grows with $N_{\rm max}$. This feature might give hope that this process can be carried out indefinitely, and that in the limit $N_{\rm max}\to\infty$ all states below the sister trajectory decouple. Yet, while this would not necessarily imply a contradiction with the theorem presented in this work, as infinitely many trajectories lie above the sister trajectory at sufficiently large energies, we provide evidence that this is not the case. Rather, the $N_{\rm max}\to \infty$ limit is ill-defined, as in that case the values of the ansatz parameters become infinite, and thus yield a non-sensical infinite value for the amplitude.

\begin{figure}[h]
    \centering
    \includegraphics[width=\textwidth]{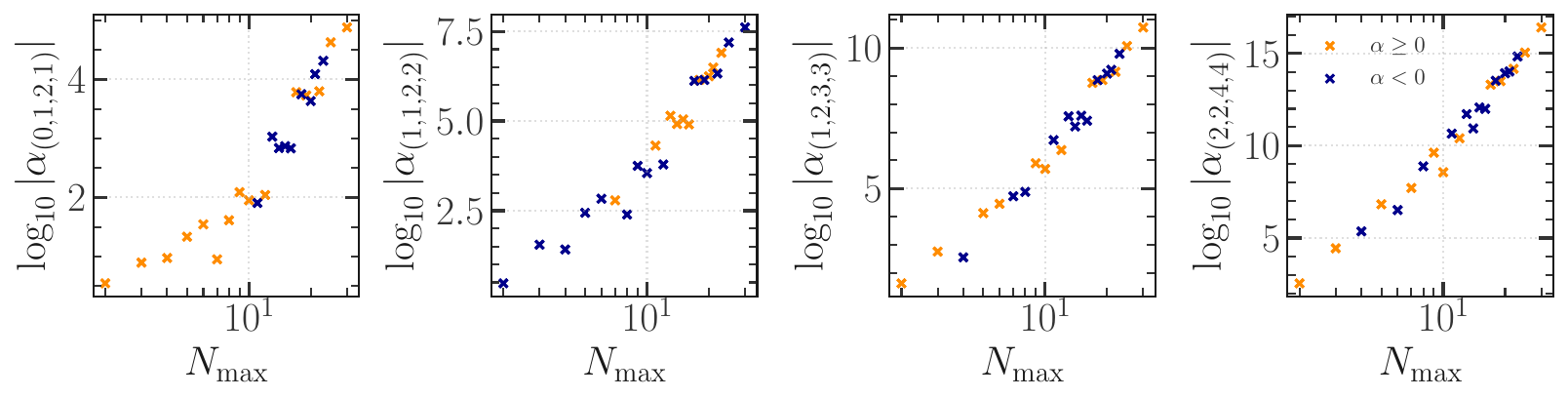}
    \caption{Here we display the evolution of the first four coefficients of the amplitude. As $N_{\rm max}$ increases, these coefficients acquire a clear, power-law-like growth, which demonstrates that the resulting amplitude does not admit a finite limit. Orange data points represent positive values, while blue points refer to negative values of the respective parameter. This is complemented with fig.~\ref{fig:amplitude-growth}, which shows the growth of the amplitude itself.}
    \label{fig:alpha-growth}
\end{figure}

We show this in fig.~\ref{fig:alpha-growth}, where we provide the evolution with $N_{\rm max}$ of various coefficients $\alpha_{(c_s,c_{tu}, d_s, d_{tu})}$ in logarithmic scale. These plots show clearly that the value of the ansatz coefficients exhibits power-law growth with $N_{\rm max}$, and therefore impede the existence of a well-defined $N_{\rm max}\to \infty$ limit, as in that case the infinite series defining the amplitude as in eq.~\eqref{eq:primal-ansatz} would diverge.

\begin{figure}
    \centering
    \includegraphics[scale=0.9]{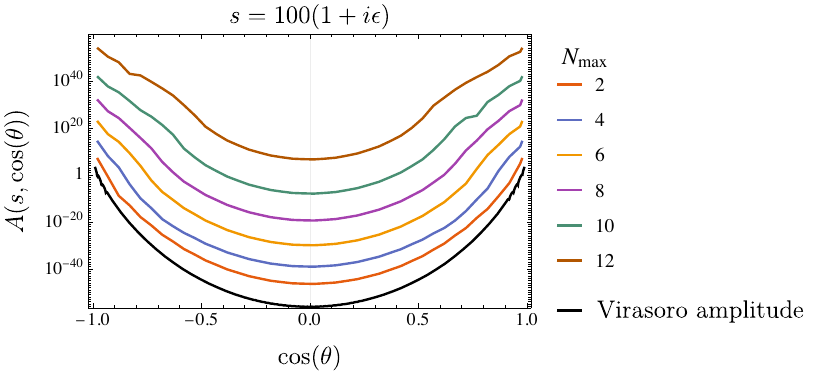}
    \caption{Plot of the amplitude itself slightly above the real axis, in the fixed angle limit, as a function of the $\cos(\theta)$. We have $s=100(1+I \epsilon)$ and $\cos(\theta) = 1+\tfrac{2t}{s}$. As $N_{\rm max}$ increases, the amplitude appears to diverge exponentially.}
    \label{fig:amplitude-growth}
\end{figure}

Relatedly, in fig.~\ref{fig:amplitude-growth} we show the growth of the extremal amplitudes themselves with $N_{\rm max}$ by plotting the value of each amplitude at fixed $s=100(1+i\epsilon)$ $\cos\theta$ between $-1$ and $1$. It is clear that the values of the amplitude grow exponentially with $N_{\rm max}$, with more than 40 orders of magnitude of difference between the Virasoro amplitude (which corresponds to $N_{\rm max}=0$) and the case $N_{\rm max}=12$. Since such behavior is obviously incompatible with a sensible $N_{\rm max}\to \infty$ limit, we conclude that it is not possible to remove all states below the sister trajectory consistently. Rather, the single-trajectory-like behavior that we observe in our numerics must always be transient, with all the couplings below the sister trajectory becoming relevant at some finite energy scale.

We can still extract some interesting physics from fig.~\eqref{fig:amplitude-growth}. 
Roughly, we expect weak-coupling to break down at energies such that the amplitude becomes of order 1. However, in our numerics we used units in which the residue of the massless pole is $8 \pi G_N =1$, thus in practice we normalize the amplitude by a factor of $(8 \pi G_N)^{-1}$. Therefore, we expect the weak-coupling hypothesis to be violated when the value of the numerical extremal amplitudes becomes of order$\sim (8 \pi G_N)^{-1}\sim M_{pl}^2 $, which is a priori an unknown parameter of the theory. Yet, since we work in units of $M_s=1$, where $M_s$ is the mass of the lightest massive string state, demanding that weak coupling is satisfied for such extremal amplitude will give a maximum value for the ratio $M_{pl}/M_{s}$. 
%


We elaborate on the interpretation of our result in the context of the species scale in quantum gravity in the discussion section.

What we learn from this exercise is that the region of the spectrum that is allowed to resemble a single trajectory and maintain weak coupling must always be confined to masses well below the Planck scale, with the full infinite set of Regge trajectories becoming visible as we approach $M_{pl}$.

\subsection{Asymptotic limits: Regge, fixed imaginary angle}
\label{sec:numerics-regge-asymptotics}

In this subsection, we study different asymptotic limits of the extremal amplitudes obtained by maximizing the coupling $c_{1,4}^{++}$.

We begin by showing the asymptotic behavior of the amplitudes for different $N_{\rm max}$ at large energy and fixed imaginary angles, namely the unphysical $s\to \infty$, $t \to \infty$, $s/t$ fixed limit. 

In~\cite{Caron-Huot:2016icg,Haring:2023zwu} it was shown that all weakly-coupled amplitudes for colored scalars (i.e. open-string-like) should not grow faster than the Veneziano amplitude in this regime, and here, we investigate whether a similar statement holds for our weakly-coupled graviton scattering amplitudes. Indeed, we observe that the asymptotic behavior of the extremal amplitudes in this regime is exactly the same as that of the Virasoro-Shapiro amplitude for all values of $N_{\rm max}$, apart from a constant global scale, already seen at fixed real angles in fig.~\ref{fig:amplitude-growth}. This hints at the existence of an equivalent "universality of the Virasoro-Shapiro amplitude" property generalizing the one shown in~\cite{Caron-Huot:2016icg,Haring:2023zwu} to the case of $stu$-symmetric scattering, which was conjectured already in the mentioned references.

\begin{figure}[h]
    \centering
    \includegraphics[width=0.7\linewidth]{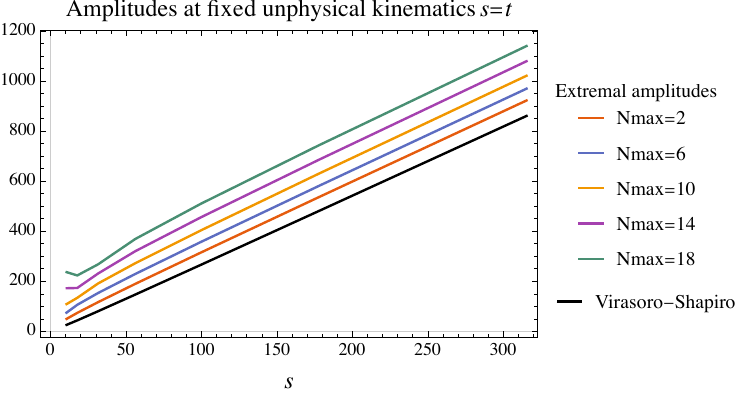}
    \caption{Asymptotic behavior of the amplitudes in the fixed (imaginary) angle limit $s\to\infty$, $s/t=1$ for various values of $N_{\rm max}$. In all cases, the growth is the same as that of the Virasoro-Shapiro amplitude, modulo an overall $N_{\rm max}$-dependent constant factor.}
    \label{fig:veneziano-asymptotics}
\end{figure}

\begin{figure}
    \centering
    \includegraphics[scale=1]{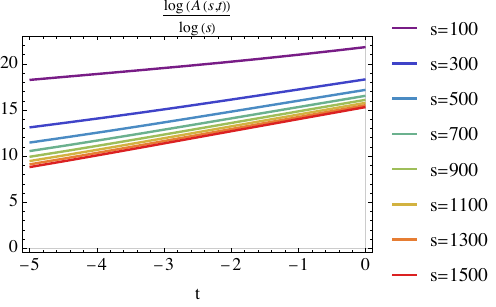}
    \caption{Ratio $\frac{\log(A_{N_{\rm max}}(s,t))}{\log(s)}$ as a function of $t$ for various values of $s$ and $N_{\rm max}=10$. We observe a very linear dependence that appears to converge at large enough $s$, showing a neat linear trajectory with constant slope, as in eq.~\eqref{eq:numericalRT}.
    }
    \label{fig:reggetrajectories}
\end{figure}

Now we turn to the Regge limit $s\to\infty$, $t<0$ to shed direct light on the emergence of the sister trajectory. Assuming that the amplitude reggeizes and therefore behaves as $s^{\alpha(t)}$ at large $s$, we can measure an effective Regge growth by considering the following quantity, at fixed $t\leq 0$:
\begin{equation}
    \lim_{s\to\infty} \frac{\log(A_{N_{\rm max}}(s,t))}{\log(s)}.
\end{equation}
If we measure this ratio at an $s$ large enough for the amplitude to have neatly reggeized, i.e.~we have
\begin{equation}
    A_{N_{\rm max}}(s,t) = f(t) s^{\alpha(t)},\quad \alpha(t) = \alpha_0+\alpha't+\mathcal{O}\left(t^2\right),
\end{equation}
then 
\begin{equation}
\label{eq:numericalRT}
     \frac{\log(A_{N_{\rm max}}(s,t))}{\log(s)} = \alpha(t) + \frac{\log(f(t))}{\log(s)},
\end{equation}
which allows to  obtain the leading trajectory $\alpha(t)$ numerically.
This is shown for $N_{\rm max}=10$ in fig.~\ref{fig:reggetrajectories}, where the ratio~\eqref{eq:numericalRT} is plotted as a function of $t$ for various values of $s$.

\begin{figure}
    \centering
    \includegraphics[scale=0.9]{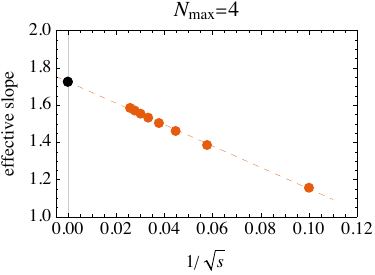}
    \includegraphics[scale=0.9]{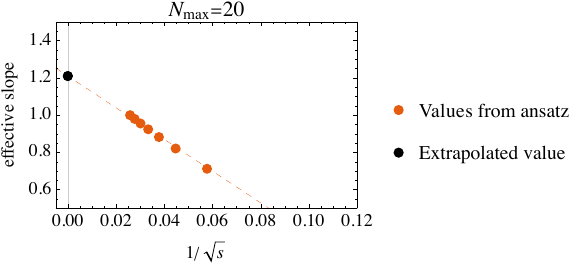}
    \caption{Typical extrapolation in $N_{\rm max}$ shows good convergence with $1/\sqrt{s}$. Every single value of $N_{\rm max}=2,\dots,25$ exhibits such a nice scaling, making it possible to extrapolate the value of the slope to infinite $s$. }
    \label{fig:RTconvergence}
\end{figure}

For all values of $N_{\rm max}$, we observe a very linear dependence on $t$, which allows for a simple extraction of the effective Regge slope $\alpha'_{\rm eff}(N_{\rm max})$. This is shown for $N_{\rm max}=4$ and 20 as well as various values of $s$ in fig.~\ref{fig:RTconvergence}, showing reasonable convergence in $s$.

Extracting the effective Regge intercept $\alpha_{0,\rm eff}(N_{\rm max})$ is more delicate, as the ratio of logarithms in~\eqref{eq:numericalRT} varies very slowly with $t$ and behaves effectively as a constant in the range of $t$ considered. Since extremely large values of $s$ are not numerically accessible, this contribution is expected to be small but not necessarily negligible, and thus potentially pollutes the constant term coming from $\alpha(t)$. With these caveats in mind, we still extract the $t$-independent piece of the curves in fig.~\ref{fig:reggetrajectories} and identify it with $\alpha_{0,\rm eff}$, even if it should be taken with a grain of salt. 

\begin{figure}
    \centering
    \includegraphics[scale=1]{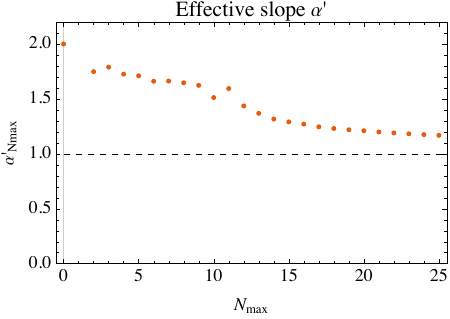}
    \caption{Effective Regge slope as a function of $N_{\rm max}$. Note that the ansatz at $N_{\rm max}=1$ is just the Virasoro-Shapiro amplitude and thus has slope 2, while at large $N_{\rm max}$ we see a transition towards the effective slope of the sister trajectory, $\alpha'_{\rm eff}\sim1$. At $N_{\rm max}=1$ no solutions are found because the ansatz does not have enough parameters and $N_{\rm max}=0$ is simply the V-S term.}
    \label{fig:slope}
\end{figure}

Finally, in fig.~\ref{fig:slope}, we show the extracted effective slope as a function of $N_{\rm max}$. It illustrates that she slope appears to neatly evolve from $\alpha'=2$ for low $N_{\rm max}$, consistent with the Virasoro-Shapiro amplitude corresponding to $N_{\rm max}=0$, towards the slope of the sister trajectory $\alpha'=1$ at large $N_{\rm max}$, showing that it is indeed the emergent trajectory that is controlling the Regge limit of the extremal amplitudes, at least in these ranges of energy.

\newpage

\paragraph{Comparison with dual optimization results \cite{Albert:2024yap,Albert:2022oes,Albert:2023jtd,Albert:2023seb}}

At large $N$~\cite{tHooft:1973alw}, gauge theories become theories of exactly stable mesons~\cite{Witten:1979kh}. Using dual bootstrap techniques, several studies have tried to locate the S-matrix of large-N gauge theories in the allowed space of amplitudes recently~\cite{Albert:2022oes,Albert:2023jtd,Albert:2023seb,Fernandez:2022kzi,Ma:2023vgc,Li:2023qzs,Huang:2025icl}. In particular, the works \cite{Albert:2022oes,Albert:2023jtd,Albert:2023seb} identified precisely extremal amplitudes with an apparent spectrum made of a single trajectory, in possible tension with the theorems of the present paper and~\cite{Eckner:2024pqt}. While we cannot exactly identify the source of the tension, the results of this section might contribute to lessening that tension.

Focussing, for instance, on \cite{Albert:2023seb}, the authors detect a large, possibly curved, Regge trajectory, see fig~8b. 
This figure also shows possibly spurious states between the curved trajectory and the diagonal $J=m^2/m_\rho^2$. Given our numerical results, we could suggest that another possible interpretation would be that what this figure shows is the emergence of a dominant \textit{sister} trajectory, with states between the sister and the leading trajectory. This picture is exactly similar to what we observe in our primal ansatz, see our figs.~\ref{fig:amplitude-main},~\ref{fig:max-alt-c}.

In gravity, which is really the subject of the numerics presented in this section anyway, the authors of~\cite{Albert:2024yap} also identified single-trajectory-like behaviors. Figs. 8, 10, of the aforementioned paper in particular also show possible residual, not fully decoupled states around a leading trajectory. Fig.~15, however, does not. Overall, the situation appears more contrived in this case.

One last element of speculation relates to hypotheses used to clean up the spectrum. By default, SDPB produces spectra with a plethora of states above the leading trajectory, see e.g.~\cite[fig.~8.a]{Albert:2024yap}. These states appear spurious and can be removed to obtain healthier-looking solutions. However, we would like to suggest that maybe in some way these ``upper-triangular'' states could correspond to the states lying between a hypothetical dominant sister and a possible leading trajectory.

Overall, apart from the solid fact that single-trajectory amplitudes cannot exist, we cannot provide a rock-solid piece of evidence to solve the tension. Nevertheless, we hope that the numerics we presented in this section can provide interesting directions to investigate in the context of this tension.


\section{Conclusion \& Discussion}
\label{sec:discussion}

In this work, we have revisited the structure of weakly coupled gravitational amplitudes from two complementary angles: an analytic derivation based on fundamental S-matrix principles and a constructive numerical analysis using a primal bootstrap.

Regarding the analytic side, we derived a no-go theorem for amplitudes with finitely many daughter trajectories. We demonstrated that any meromorphic, crossing-symmetric, and unitary graviton amplitude composed of a finite number of Regge trajectories must necessarily violate polynomial boundedness. By tracing the asymptotic decay of couplings required by crossing symmetry, we showed that such spectra enforce an exponential growth of the amplitude at high energies, thereby violating causality. This result simplifies and generalizes previous arguments for $s,t$-symmetric amplitudes. 
Our arguments, therefore, imply that a classical rotating string needs to be able to support excitations in order to be compatible with relativistic interactions at weak coupling. A trivial related comment is that, with usual strings, in order to get rid of excitations, the tension should be taken to infinity, in which case the string reduces to a point-like object, not a reggeized, extended object.

On the numerical front, we investigated how ``single-trajectory-like'' solutions can nonetheless appear to emerge as extremal configurations in the bootstrap. Using a primal bootstrap approach based on the Häring-Zhiboedov ansatz, we observed that maximizing the leading coupling drives the system towards a specific spectrum structure characterized by a dominant \textit{sister} trajectory with halved slope, $j_{\rm sister}(t)\sim t$, independently of the couplings being extremized. Our analysis further revealed that these extremal solutions are spurious and an artefact of a truncated ansatz. While the ansatz effectively decouples subleading states below the sister trajectory, the amplitude itself diverges as the ansatz size goes to infinity, making it unphysical. In addition, states between the numerically dominant sister trajectory and the leading trajectory do not decouple in our construction, raising the question of whether further constraints could be imposed.

\paragraph{Weak coupling and resolvability of the string}
In the text, we also discussed our relation to weak coupling. What we saw is that extremal amplitudes look ``less stringy'' and push the onset of daughter trajectories, hence the scale at which one can resolve the string fully, in the UV. But this goes together with generating exponentially large amplitudes compared to gravitational scattering, which suggests that the theory might want to become strongly coupled. Therefore, our results imply that it would be possible for a string to not be resolvable, only if it is strongly coupled. At weak coupling, all the string-scale excitations of the string need to be visible and come in large numbers. It would be interesting to deepen these points in relation to~\cite{Haring:2024wyz,Caron-Huot:2024lbf} and the emergence of the species scale~\cite{Dvali:2007hz,Dvali:2009ks,Dvali:2010vm,Dvali:2012uq} in quantum gravity.

\paragraph{Understanding the role of the sister trajectory.}A natural question that would be interesting to understand is the reason why sister trajectories emerge when extremizing the couplings, and related, whether we can put more rigorous bounds on the energy scale where subleading trajectories should kick in. A possible avenue to increase our understanding of the single-trajectory-like amplitudes and role of the sister would be to study a sort of reciprocal primal problem, by demanding that the coefficients below a given trajectory (the leading one or some daughter/sister) and without extremizing any coefficients, and seeing if the resulting solutions happen to saturate the bounds on some couplings.

\paragraph{Extending the numerical approach to $d=10$ supergravity.} Another interesting exercise would be to extend our numerical analysis to the scattering of scalar particles in the graviton supermultiplet in $d=10$ supergravity. This would allow us to make the connection to~\cite{Albert:2024yap} more explicit, and check whether our primal construction still generates single-trajectory-like extremal amplitudes in that case. Going the other way, in order to make contact with large $N$ gauge theories, it would be interesting to study the open-string version of the stringy ansatz, to see if there as well, an emergent sister dominates the scattering. There, the problem, already identified in~\cite{Haring:2023zwu}, is that convergence cannot be reached at finite ansatz sizes and a double scaling limit $N_{\rm max},N_{\rm pole}\to\infty$ should be taken, which renders the numerics more cumbersome.

\paragraph{Elucidating the nature of the stringy excitations.}
The results in this work show that the spectrum of a UV-complete, \textit{weakly coupled} theory of interacting gravitons must exhibit very stringy features, with infinitely many massive higher spin states organized into an infinite tower of Regge trajectories. We know from the Weinberg-Witten theorem that the graviton cannot be a composite state~\cite{Weinberg:1980kq}. It would be interesting to be able to take one more step and rigorously prove that the stringy UV completion is a genuine theory of fundamental strings, rather than some string-like excitation of another kind of matter. 

Such a proof would surely entail studying higher-point scattering, which allows in particular to probe the growth of the density of states in relation to the daughters as observed in~\cite{Gross:1969db} long ago, see also \cite{Bardakci:1969asw,Fubini:1969wp}. Recent progress has been achieved for studying dual models at higher-multiplicity~\cite{Cheung:2023uwn,Arkani-Hamed:2023jwn,Komatsu:2025onf} as well as higher-point Regge limits~\cite{Costa:2023wfz}, and it would be interesting to build on these works.

\paragraph{Coulomb branch of $\mathcal{N}=4$ super-Yang-Mills.} The argument we presented is severe enough that it cannot be cured easily. For instance, adding finitely many loops in the UV can never give rise to poles in the cross-channel. Therefore, the crossed-poles would still have to come from the direct channel poles, giving rise to the same inconsistency that we describe.

There is one case, however, recently studied in the literature~\cite{Alday:2025pmg}, when finitely many trajectories of \textit{physical} bound states, it is the case of the Coulomb branch of $\mathcal{N}=4$ super-Yang-Mills at large $N_c$. It was observed that in this case, while infinitely many trajectories clearly exist, at finite 't Hooft coupling, only finitely many of them give rise to physical bound states below some ionisation threshold.  Basically, what happens in this case is that below a critical coupling, the string is too floppy to support physical quantum excitations before the threshold, see also previous discussions in \cite{Klebanov:2006jj,Maldacena:2022ckr}.\footnote{We thank S. Zhiboedov for a discussion on this point} Furthermore, the threshold physics is very stringy: it corresponds to infinitely many gluons being exchanged, building up the full flux-tube.
At infinite 't Hooft coupling, the amplitude reduces to the Veneziano amplitude, which is fully meromorphic with infinitely many trajectories. 

Overall, many physical elements make this $\mathcal{N}=4$ construction depart from our axioms and no strict meromorphic weakly coupled amplitude exists with finitely many trajectory in the sense we describe in the paper. Nevertheless, it would be very interesting to study the connection between this construction and our argument further.

\hrulefill

\paragraph{Acknowledgements}

We would like to thank Justin Berman, Nick Geiser, Mehmet Gumus, Kelian Häring (especially for sharing a mathematica notebook with code), Yu-tin Huang, Xi Yin and Sasha Zhiboedov for interesting discussions and comments. In particular we would like to thank Mehmet Gumus for earlier collaboration on a related question, Sasha Zhiboedov for numerous discussions and suggesting to look at primal constructions using their ansatz~\cite{Haring:2023zwu} to shed another light on our study~\cite{Eckner:2024pqt} in the context of the study of matching with~\cite{Albert:2022oes,Albert:2023jtd,Albert:2023seb,Albert:2024yap}, Xi Yin for discussions on the meaning of subleading Regge trajectories, and Justin Berman for thorough comments on the draft.

The work of PT is supported by Agence Nationale de la Recherche (ANR), project ANR-22-CE31-0017.

The work of FF is supported by the European Research Council (ERC) under the European Union’s Horizon 2020 research and innovation programme (grant agreement No 101002551).

This publication has received funding from the European Union’s Horizon Europe research and innovation program under the Marie Sk\l odowska-Curie COFUND Postdoctoral Programme grant agreement No.101081355-SMASH and from the Republic of Slovenia and the European Union from the European Regional Development Fund. \textbf{Disclaimer}: Co-funded by the European Union. Views and opinions expressed are, however, those of the authors only and do not necessarily reflect those of the European Union or European Research Executive Agency. Neither the European Union nor the granting authority can be held responsible for them.

\appendix

\section{Wigner polynomials and their asymptotic behavior}
\label{app:Wigner}

Wigner polynomials, sometimes called Wigner D-matrices, are the generalization of the Legendre polynomials signaling the exchange of a spinning state at a pole in the scattering of external scalars to the case in which the external states carry helicity. They can be written in terms of hypergeometric functions as
\begin{equation}
\begin{split}
  d^J_{\lambda \lambda'}(z)=\left(\frac{1+z}{2}\right)^{\frac{\lambda'+\lambda }{2}}  \left(\frac{1-z}{2}\right)^{\frac{\lambda'-\lambda }{2}} \sqrt{\frac{(J-\lambda )! (J+\lambda')!}{(J+\lambda )!
   (J-\lambda')!}} \,\\ \times \, _2F_1\left(\lambda'-J,J+\lambda'+1;-\lambda +\lambda'+1;\frac{1-z}{2}\right),   
\end{split}
\end{equation}
where $z=\cos\theta$.
Their asymptotic behavior at large $J$ is needed to obtain a constraint on the decay of the $c_{nJ}$ coefficients at large $n$ from requiring that the single-trajectory amplitude generates the required poles in the crossed channel, and is given (in terms of $\theta$) by\footnote{See for instance~\cite{Hoffmann_2018}}
\begin{equation}
\begin{split}
   d^J_{\lambda \lambda'}(\theta)\sim(-1)^{\lambda'-\lambda} & \sqrt{\frac{(J-\lambda )! (J+\lambda')!}{(J+\lambda )!
   (J-\lambda')!}}\frac{1}{\Delta(J,\lambda',\lambda)^{\lambda'-\lambda}}\\
   \times & \left(\frac{\theta}{\sin \theta}\right)^{1/2}J_{\lambda'-\lambda}\left(\Delta(J,\lambda',\lambda)\theta\right),
\end{split}
\end{equation}
where $J_{\lambda'-\lambda} $ is a Bessel $J$-function and 
\begin{equation}
    \Delta(J,\lambda',\lambda)=\sqrt{J(J+1)-\frac{1}{3}(\lambda'^2+\lambda^2+\lambda'\lambda-1)}.
\end{equation}

Note that in the case that we are interested in, $J \gg \lambda, \lambda'$, in which case we have
\begin{equation}
\sqrt{\frac{(J-\lambda )! (J+\lambda')!}{(J+\lambda )!
   (J-\lambda')!}}\frac{1}{\Delta(J,\lambda',\lambda)^{\lambda'-\lambda}}=1+\mathcal{O}\left(\frac{\lambda'}{J},\frac{\lambda}{J}\right),
\end{equation}
and thus
\begin{equation}
   d^J_{\lambda \lambda'}(\theta)\sim(-1)^{\lambda'-\lambda}  \left(\frac{\theta}{\sin \theta}\right)^{1/2}J_{\lambda'-\lambda}\left(J\theta\right),
\end{equation}
where we used that $\Delta(J,\lambda',\lambda)\sim J$ in this regime. 

At large values of their argument, the Bessel functions behave as
\begin{equation}
    J_{\nu}(x) \sim \sqrt{\frac{2}{\pi x}}\cos\left(x-\frac{\nu \pi}{2} -\frac{\pi}{4}\right), \quad x \to \infty.
\end{equation}

To write these asymptotic expressions in terms of $z=\cos\theta$ we simply expand $\arccos z$ around $z=1$, as the terms we will care about come from the large mass tail of the sum over poles in the amplitude, and therefore $z_n=1+\frac{2t}{m_n^2}\to1$. Since $\arccos z = i \sqrt{2(z-1)} +\mathcal{O}((z-1)^{3/2})$, plugging this back into the asymptotic expansions of the relevant Bessel function we get
\begin{equation}
\label{eq:Wignerdasymptotic}
   d_{\lambda\lambda'}^J(z)\sim(-1)^{\lambda'-\lambda}\sqrt{\frac{1}{\pi J }}\left(\frac{2}{z^2-1}\right)^{1/4}\cos\left(iJ \sqrt{2(z-1)}-\frac{J \pi}{2} \right) \sim e^{J\sqrt{2(z-1)}}.
\end{equation}

Note that the exponential behavior is independent of the helicities of the external particles. Moreover, replacing $z=1+\frac{2t}{m_n^2}$ we find the exponential growth $d^J_{\lambda \lambda'}\sim e^{2J\sqrt{t/m_n^2}}$.

\section{Alternative proof in terms of partial wave coefficients}

To make contact with the proof of the need for infinitely many Regge trajectories in the case of massive scalar external states presented in~\cite{Eckner:2024pqt}, we provide an alternative proof for the graviton case following the same logic, namely computing the partial wave coefficients via the Froissart-Gribov projection and showing that they are ill-defined. 

It is well known that in four dimensions, infrared divergences impede the straightforward usage of the partial wave expansion in theories with massless mediators such as gravity. The reason for this is that the long-range nature of the interaction causes the partial wave coefficients $f_J(s)$ to diverge. 

Consider, for example, the tree-level scattering of scalar particles involving gravitational interactions:
\begin{equation}
    \mathcal{M}\left(s,t\right)=-\frac{8 \pi G_N s^2}{t}+ \, \text{regular as $t \to 0$}
\end{equation}In this case, the partial waves are given by
\begin{equation}
\label{eq:examplePWE}
    f_J(s)=\mathcal{N}_d\int_{-1}^1 {\rm d}z \, (1-z^2)^\frac{d-4}{2} \mathcal{P}_J^{(d)}(z) \, \mathcal{M}\left(s,t(z,s)\right) = \mathcal{N}_d \int_{-1}^1  {\rm d}z \, (1-z^2)^\frac{d-4}{2}\frac{16 \pi G_N s}{1-z}+ \dots,
\end{equation}
where we isolated the universal contribution from the graviton pole and used $z=1+2t/s$, and where $\mathcal{N}_d=\frac{(16 \pi)^\frac{2-d}{2}}{\Gamma(\frac{d-2}{2})}$. 

It is clear from eq.~\eqref{eq:examplePWE} that the contribution to the partial wave coefficients coming from the graviton pole has a logarithmic divergence in $d=4$, which precisely corresponds to the fact that gravitational interactions do not decay fast enough for gravitons to be well-defined asymptotic states in four dimensions. However, this IR divergence is universal to all gravitational theories and can be dealt with systematically in various ways, such as using dimensional regularization or considering a fictitious mass for the graviton as an IR regulator (see for instance \cite{Weinberg:1965nx,Oller:2022ozd,Ware:2013zja}). In particular, this divergence has nothing to do with whether the UV completion contains a single or infinitely many Regge trajectories, and therefore it is reasonable for our purposes to extract the IR divergent piece and use the finite part of the partial wave coefficients to check the consistency of our candidate single-trajectory gravitational amplitude.

We define
\begin{equation}
    f_J(s)=f_J^{\text{pole}}(s) + \tilde{f}_J(s),
\end{equation}
where $f_J^{\text{pole}}(s)$ as the (IR divergent) contribution to the partial wave coming from the $1/t$ pole and $\tilde{f}_J$ is the part sensitive to the UV completion. The pole contribution can be computed in closed form in dimensional regularization, and we can focus on the rest as a probe for the consistency of the amplitude. Concretely, the contribution from the graviton pole can be obtained by using
\begin{equation}
\begin{split}
   \int_{-1}^1 {\rm d}z \, (1-z^2)^\frac{d-4}{2} \frac{\mathcal{P}_J(z)}{1-z}=& -\frac{\sqrt{\pi } 2^{1-\frac{J}{2}} (J-1)\text{!!} \Gamma \left(\frac{d-4}{2}\right)
   \left(3-\frac{d}{2}\right)_{\frac{J}{2}}}{\frac{J}{2}! \Gamma \left(\frac{1}{2} (d+J-3)\right)} \quad \text{for $J$ even}, \\=&-\frac{\sqrt{\pi } 2^{\frac{1}{2}-\frac{J}{2}} J\text{!!} \Gamma \left(\frac{d-4}{2}\right)
   \left(3-\frac{d}{2}\right)_{\frac{J-1}{2}}}{\Gamma \left(\frac{J+1}{2}\right) \Gamma
   \left(\frac{1}{2} (d+J-2)\right)} \quad \text{for $J$ odd},
\end{split} 
\end{equation}
where $(x)_{n}=x (x+1)\dots (x+n-1)$. The IR divergent piece can be extracted by taking the $d \to 4+ \epsilon $ limit, and is given by
\begin{equation}
    f_J^{\text{pole}}(s)=-\frac{(16 \pi)^2 G_N s}{\epsilon}+ \, \text{finite},
\end{equation}
and is remarkably independent of $J$.

Therefore, IR divergences are a universal characteristic of 4-dimensional gravitational theories coming purely from having GR as the EFT valid at long distances, and do not say anything about the consistency of a possible UV completion. Moreover, this divergence can be regularized in a systematic, well-understood way, to give a finite, universal contribution to the partial wave coefficients.

For these reasons, we focus from now on on the contribution to the partial wave coefficients coming from the massive states, by defining
\begin{equation}
    \tilde{T}^J_{\lambda_1\lambda_2\lambda_3\lambda_4}(s)=T^J_{\lambda_1\lambda_2\lambda_3\lambda_4}(s)-T^{J \text{\,pole}}_{\lambda_1\lambda_2\lambda_3\lambda_4}(s),
\end{equation}
where $T^{J \text{\,pole}}_{\lambda_1\lambda_2\lambda_3\lambda_4}(s)$ is the (IR divergent) contribution to the partial wave coming from the $1/t$ pole and $\tilde{T}^J_{\lambda_1\lambda_2\lambda_3\lambda_4}(s)$ is the part sensitive to the UV completion. 

The point is that $\tilde{T}^J_{\lambda_1\lambda_2\lambda_3\lambda_4}(s)$ should be finite for any sensible UV completion having general relativity as a low-energy limit. We will show now that this is not the case if the UV consists of a single trajectory.

To compute the partial wave coefficients we use the generalization to spinning states of the Froissart-Gribov projection. The idea is the same as in the spinless case, and is based on the existence of a function $e^J_{\lambda \mu}(z)$ whose discontinuity is related to $d^J_{\lambda \mu}(z)$.

More precisely, define
\begin{equation}
    \label{eq:Wignere}
    \begin{split}
        &e^J_{\lambda \mu}(z)=\frac{(-1)^{\lambda-\mu}}{2}[\Gamma(J+\lambda+1)\Gamma(J-\lambda+1)\Gamma(J+\mu+1)\Gamma(J-\mu+1)]^{1/2}\left(\frac{1+z}{2}\right)^{\frac{\lambda+\mu}{2}}\\
        & \times \left(\frac{1-z}{2}\right)^{-\frac{\lambda-\mu}{2}}\left(\frac{z-1}{2}\right)^{-J-\mu-1}\frac{1}{\Gamma(2J+2)} {}_2F_1\left(J+\lambda+1,J+\mu+1,2J+2;\frac{2}{z-1} \right),
    \end{split}
\end{equation}
which satisfies
\begin{equation}
    e^J_{\lambda \mu}(z+i\epsilon)-e^J_{\lambda \mu}(z-i \epsilon)=-i \pi \, d^J_{\lambda \mu}(z), \quad \text{for \,} z \in (-1,1).
\end{equation}

This allows us to express the projection onto a given partial wave coefficient as a contour integral:
\begin{equation}
\label{eq:preFG}
    T^J_{\lambda_1\lambda_2\lambda_3\lambda_4}(s)=\oint_{\mathcal{C}} {\rm d}z  \, d^J_{\lambda_{12},\lambda_{34}}(z) \, \mathcal{M}_{\lambda_1\lambda_2\lambda_3\lambda_4}(s,t(z)), 
\end{equation}
where $\mathcal{C}$ is a contour around the segment $(-1,1)$. 

Blowing up the contour, we obtain the Froissart-Gribov projection for the partial wave coefficients, which expresses them in terms of the discontinuities of the amplitude:
\begin{equation}
\label{eq:FG}
\begin{split}
    T^J_{\lambda_1\lambda_2\lambda_3\lambda_4}(s)=\frac{1}{i \pi}&\bigg( \int_{z_R}^\infty {\rm d}z  \, e^J_{\lambda_{12},\lambda_{34}}(z) \, {\rm Disc}\mathcal{M}_{\lambda_1\lambda_2\lambda_3\lambda_4}(s,t(z))\\
    &+\int_{-\infty}^{z_L} {\rm d}z  \, e^J_{\lambda_{12},\lambda_{34}}(z) \, {\rm Disc}\mathcal{M}_{\lambda_1\lambda_2\lambda_3\lambda_4}(s,t(z))\bigg),
\end{split}
\end{equation}
where $z_L, z_R$ are the values of $z$ corresponding to the leading singularities in the $u$ and $t$ channels. Note that we are neglecting massless poles that would also enter this equation because we will apply this formula to $\tilde{T}^J_{\lambda_1\lambda_2\lambda_3\lambda_4}(s)$, which is finite at $t=0$, $u=0$.

An important remark is that the Froissart-Gribov projection is only valid if we can legitimately neglect the arch at infinity when blowing up the contour in~\eqref{eq:preFG}. Since the large $z$ behavior of the $e^J_{\lambda \mu}(z)$ functions is $e^J_{\lambda \mu}(z) \sim z^{-J-1}$ as $z\to \infty$, if the amplitude behaves as $\mathcal{M}_{\lambda_1\lambda_2\lambda_3\lambda_4}(s,t(z)) \sim z^{\alpha(s)}$ at large $z$, eq.~\eqref{eq:FG} is valid for $J>\alpha(s)$.

Let us assume that we are in the regime of validity of the Froissart-Gribov projection and use it to relate the partial wave coefficients to the couplings between the massive states and the graviton.

Consider first the $++--$ helicity configuration. Writing $z$ in terms of $t$ we have:
\begin{equation}
    \label{eq:FG++}
\begin{split}
    \tilde{T}^J_{++--}(s)&=\frac{2}{i \pi s}\bigg( \int_{m_1^2}^\infty {\rm d}t  \, e^J_{00}(1+\frac{2t}{s}) \, {\rm Disc}\mathcal{M}_{++--}(s,t,-s-t)\\
    &+\int_{-\infty}^{-m_1^2-s} {\rm d}t  \, e^J_{00}(1+\frac{2t}{s}) \, {\rm Disc}\mathcal{M}_{++--}(s,t,-s-t)\bigg),
\end{split}
\end{equation}

The contribution from the right cut can be computed using the fact that crossing relates $\mathcal{M}_{++--}(s,t,u)$ to $\mathcal{M}_{+-+-}(t,s,u)$, and thus
\begin{equation}
\begin{split}
    \int_{m_1^2}^\infty {\rm d}t &  \, e^J_{00}(1+\frac{2t}{s}) \, {\rm Disc}\mathcal{M}_{++--}(s,t,-s-t)=\int_{m_1^2}^\infty {\rm d}t  \, e^J_{00}(1+\frac{2t}{s}) \, {\rm Disc}\mathcal{M}_{+-+-}(t,s,-s-t)\\
   & = -\pi \sum_{n=1}^\infty c^{+-}_{n,2n+2} \, e^J_{00}(1+\frac{2m_n^2}{s}) \, d^J_{4,-4}(1+\frac{2s}{m_n^2}).
\end{split}
\end{equation}

The left cut can be evaluated with the same logic, using also that $\mathcal{M}_{++--}(s,t,u)$ is symmetric under $t \leftrightarrow u$, and changing integration variables to $u$:
\begin{equation}
\begin{split}
    &\int_{-\infty}^{-m_1^2-s} {\rm d}t  \, e^J_{00}(1+\frac{2t}{s}) \, {\rm Disc}\mathcal{M}_{++--}(s,t,-s-t)\\ &=\int_{m_1^2}^{\infty} {\rm d}u   \, e^J_{00}(-1-\frac{2u}{s}) \,(-) {\rm Disc}\underbrace{\mathcal{M}_{++--}(s,u,t)}_{=\mathcal{M}_{+-+-}(u,t,-u-t)}.
\end{split}
\end{equation}
Note the crucial minus sign coming from the fact that ${\rm Disc}_t \to -{\rm Disc}_u $ when changing variables from $t$ to $u$.

The $e^J_{\lambda \mu}(z)$ satisfy the reflection property
\begin{equation}
    \label{eq:Wignerereflection}
    e^J_{\lambda \mu}(-z)=(-1)^{J+1+\lambda-2\mu} e^J_{\lambda,- \mu}(z).
\end{equation}
Applying this to the left cut, and noting that the integrand is now the same as for the right cut, we have
\begin{equation}
\begin{split}
    &\int_{-\infty}^{-m_1^2-s} {\rm d}t  \, e^J_{00}(1+\frac{2t}{s}) \, {\rm Disc}\mathcal{M}_{++--}(s,t,-s-t)=\\&(-1)^J\int_{m_1^2}^\infty {\rm d}u  \, e^J_{00}(1+\frac{2u}{s}) \, {\rm Disc}\mathcal{M}_{+-+-}(u,s,-s-u)=\\
   &  -\pi (-1)^J \sum_{n=1}^\infty c^{+-}_{n,2n+2} \, e^J_{00}(1+\frac{2m_n^2}{s}) \, d^J_{4,-4}(1+\frac{2s}{m_n^2}).
\end{split}
\end{equation}

Summing the contributions from both branch cuts, we obtain
\begin{equation}
\label{eq:TJ++}
    \tilde{T}^J_{++--}(s)=\frac{2i}{s}(1+(-1)^J)\sum_{n=1}^\infty c^{+-}_{n,2n+2} \, e^J_{00}(1+\frac{2m_n^2}{s}) \, d^J_{4,-4}(1+\frac{2s}{m_n^2}).
\end{equation}

Note that this partial wave coefficient depends only on the $c^{+-}$ coefficients. To obtain a set of partial wave coefficients that depends on the $c^{++}$ we can look at a process in a different channel. For instance, consider the partial wave decomposition of the $t$-channel process $\mathcal{M}_{+-+-}(t,s,u)$:

\begin{equation}
    \label{eq:FG+-}
\begin{split}
    \tilde{T}^J_{+-+-}(t)&=\frac{2}{i \pi t}\bigg( \int_{m_1^2}^\infty {\rm d}s  \, e^J_{4,-4}(1+\frac{2s}{t}) \, {\rm Disc}\mathcal{M}_{+-+-}(t,s,-s-t)\\
    &+\int_{-\infty}^{-m_1^2-t} {\rm d}s  \, e^J_{4,-4}(1+\frac{2s}{t}) \, {\rm Disc}\mathcal{M}_{+-+-}(t,s,-s-t)\bigg),
\end{split}
\end{equation}
where we used that in the $t$-channel $\cos \theta = 1 +2s/t$.

This case can be worked out in the same way as the previous one, relating $\mathcal{M}_{+-+-}(t,s,u)$ to $\mathcal{M}_{++--}(s,t,u)$ by crossing. The only difference is that when evaluating the contribution from the left cut, changing variables from $s$ to $u$ generates the factor $e^J_{4,-4}(-1-\frac{2u}{t})$ in the integrand. Using the reflection formula for the Wigner $e$ functions, we see that this term is mapped to $(-1)^{J+1}e^J_{4,4}(1+\frac{2u}{t})$, different from $e^J_{4,-4}$. This causes odd $J$ partial wave coefficients to be non-vanishing in this channel, unlike the $s$-channel process. This is not a contradiction with the $s \leftrightarrow u$ symmetry of the $t$-channel amplitude, as unlike Legendre polynomials, Wigner polynomials with generic helicities do not have definite parity as functions of $\cos \theta$. 

The result for $\tilde{T}^J_{+-+-}(t)$ is given by
\begin{equation}
    \label{eq:TJ+-}
    \tilde{T}^J_{+-+-}(t)=\frac{2i}{t}\sum_{n=1}^\infty \left( e^J_{4,-4}(1+\frac{2m_n^2}{t})+(-1)^J e^J_{4,4}(1+\frac{2m_n^2}{t} ) \right) \,c^{+-}_{n,2n+2} \, d^J_{0,0}(1+\frac{2t}{m_n^2}).
\end{equation}

We now have all the ingredients we need to kill the amplitudes. Recall that demanding the presence of the pole at $t=m_1^2$ in~\eqref{eq:DispRelsingletraj} implied that at least one of the families of coefficients $ \{c^{++}_{n,2n+2}\}$, $\{c^{+-}_{n,2n+2}\}$, decayed with $n$ as $\exp(-4n\sqrt{m_1^2/m_n^2})$, while the other set of coefficients could in principle decay faster.

Assume that the $c^{+-}_{n,2n+2}$ are the ones with the fixed decay. Inserting this into~\eqref{eq:TJ++} together with the asymptotic behavior of the $d$ and $e$ functions, we see that the large $n$ tail of the sum defining $\tilde{T}^J_{++--}(s)$ behaves as  
\begin{equation}
\begin{split}
    \tilde{T}^J_{++--}(s)=\frac{2i}{s}(1+(-1)^J)\sum_{n=1}^\infty & \underbrace{c^{+-}_{n,2n+2}}_{\sim e^{-4n\sqrt{m_1^2/m_n^2}}} \, \underbrace{e^J_{00}(1+\frac{2m_n^2}{s})}_{\sim \left(\frac{s}{2m_n^2}\right)^{J+1}} \, \underbrace{d^J_{4,-4}(1+\frac{2s}{m_n^2})}_{\sim e^{4n\sqrt{s/m_n^2}} } \\
    &\sim \sum_n e^{4n (\sqrt{s/m_n^2}-\sqrt{m_1^2/m_n^2})}, 
\end{split}
\end{equation}
which yields a divergent series for {\bf any} $s>m_1^2$ and {\bf any} $J$.

If on the other hand we assume that it is the $c^{++}_{n,2n+2}$ set of couplings the one whose asymptotics is fixed by~\eqref{eq:asymptoticdecay}, using the same logic we see from eq.~\eqref{eq:TJ+-} that the $\tilde{T}^J_{+-+-}(t)$ are divergent for {\bf any} $t>m_1^2$ and {\bf any} $J$. 

This shows that the partial wave expansion for the putative single-trajectory amplitude is ill-defined, as even upon subtracting the graviton pole, the contribution from the massive states is infinite. 

\section{Tracing the maximally allowed space spanned by the ansatz}
\label{app:almond}

\begin{figure}[H]
    \centering
    \includegraphics[width=0.83\textwidth]{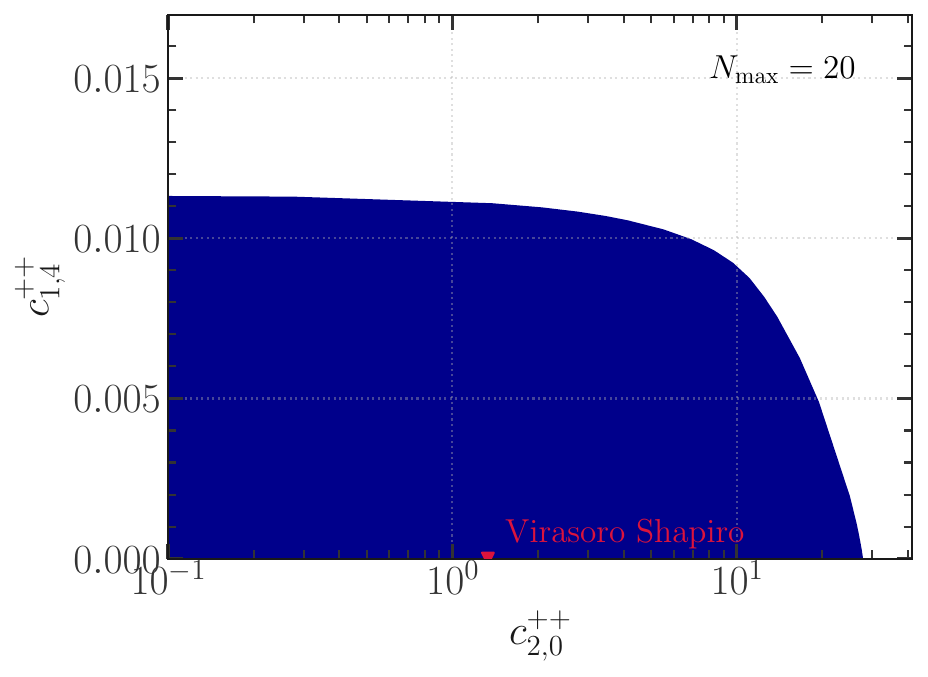}
    \caption{The allowed region for the space of the coupling tuple $(c_{1,4}^{++}, c_{2,0}^{++})$ at $N_{\rm max}=20$. Everything in blue yields an admissible pair of couplings. As a red triangle, we show the value that the Virasoro-Shapiro amplitude attains in this space.}
    \label{fig:almond}
\end{figure}

In fig.~\ref{fig:almond} we show the admissible space in the $(c_{1,4}^{++}, c_{2,0}^{++})$-plane adhering to our primal bootstrap constraints at $N_{\rm max} = 20$. It is derived by first mini- and maximizing $c_{2,0}^{++}$ and subsequently mini- and maximizing $c_{1,4}^{++}$ in the obtained extremal range. We mark the value of the Virasoro-Shapiro amplitude with a red triangle, which happens to sit on the boundary of the maximally allowed space. 

We observe for this state pair that pushing $c_{2,0}^{++}$ to its maximum forces $c_{1,4}^{++}$ to vanish. This happens in a rather sharp drop. The converse is true when the roles of $c_{2,0}^{++}$ and $c_{1,4}^{++}$ are exchanged. This anticorrelation provides a precise visualization of the decoupling phenomenon observed in, e.g., fig.~\ref{fig:amplitude-main}: maximizing the projection onto the leading trajectory ($c_{1,4}^{++}$) essentially clears out the ``lower triangle'' of the spectrum, forcing the deep subleading scalar mode ($c_{2,0}^{++}$ but also many of the other more massive scalar modes) to vanish. 

\section{Evolution of the amplitude with increasing $N_{\rm max}$}
\label{app:ampl-gig}

\begin{figure}[H]
    \centering
    \includegraphics[width=0.83\textwidth]{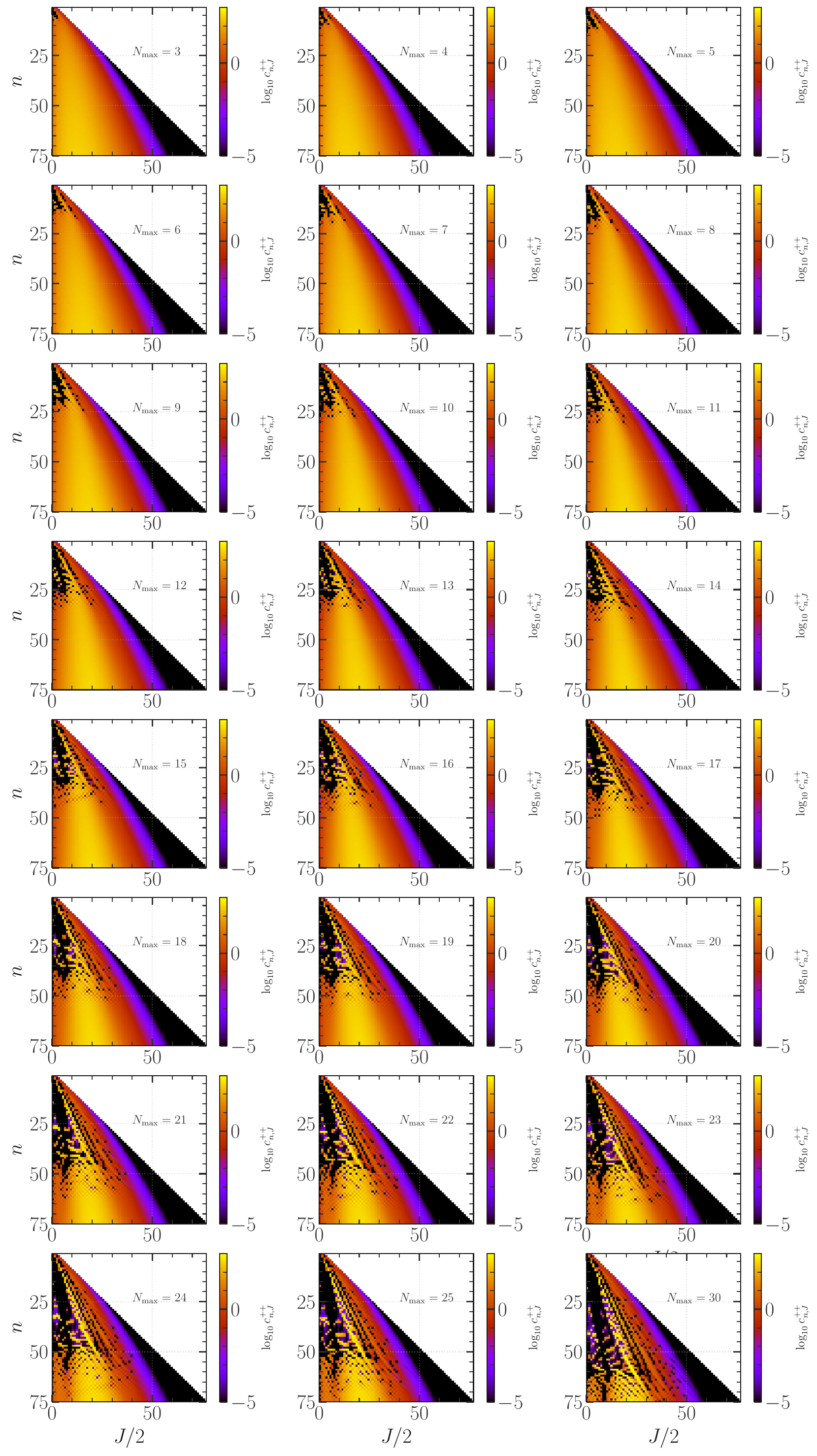}
    \caption{Evolution of the coupling data $\log_{10}c_{n,J}^{++}$ in the $(J/2,n)$-plane as the ansatz size increases from $N_{\rm max}=3$ to $N_{\rm max}=30$. The panels display heatmaps obtained by maximizing the leading coefficient $c_{1,4}^{++}$. The upper diagonal boundary corresponds to the leading Regge trajectory.}
    \label{fig:couplings-gif}
\end{figure}

In this appendix, we provide a comprehensive visualization of how the spectrum of the extremal amplitude evolves as we increase the size of our ansatz. Fig~\ref{fig:couplings-gif} displays the heatmaps of the couplings $\log_{10}c_{n,J}^{++}$ in the  $(J/2,n)$-plane, obtained by maximizing the leading coefficient $c_{1,4}^{++}$ for ansatz sizes ranging from $N_{\rm max}=3$ to $N_{\rm max}=30$.

The series of plots shows a pronounced narrowing of the spectral features as the number of parameters is increased. For small $N_{\rm max}$, the couplings exhibit a relatively wide distribution, while at large $N_{\rm max}$ the solution exhibits an emergent sister trajectory, inclined relative to the leading Regge trajectory. Most importantly, the plots indicate a gradual suppression of the couplings associated with states lying below this sister trajectory (i.e., in the lower-left portion of the spectral triangle).

\begin{figure}[H]
    \centering
    \includegraphics[width=0.83\textwidth]{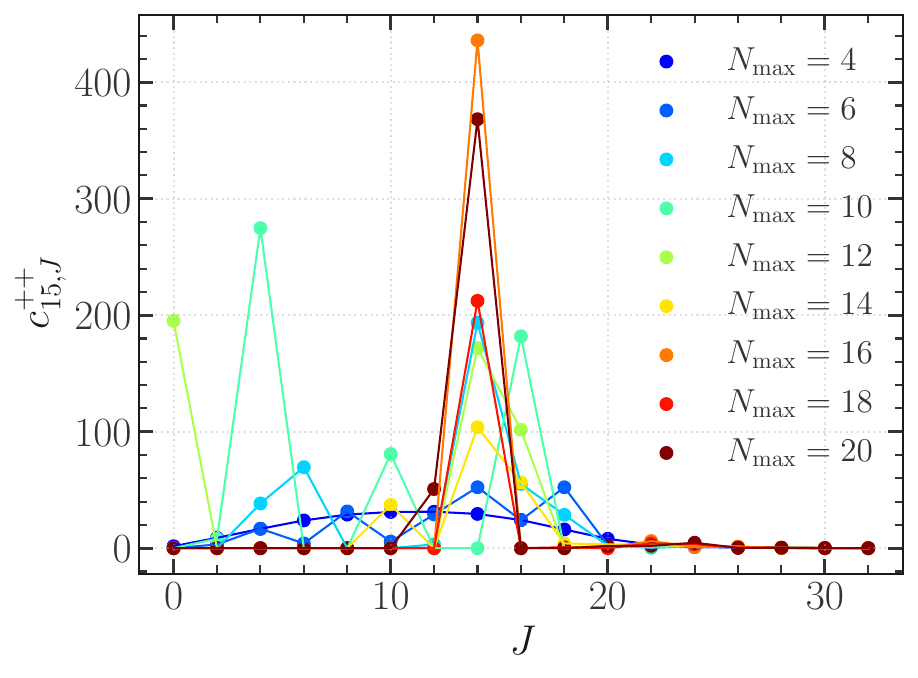}
    \caption{Same as the right panel of fig.~\ref{fig:amplitude-main} for increasing values of $N_{\rm max}\leq 20$.}
    \label{fig:horizontal-slices}
\end{figure}

We complement the right panel of fig.~\ref{fig:amplitude-main} with a plot of the same horizontal slice through the coupling matrix for increasing values of $N_{\rm max}$ in fig.~\ref{fig:horizontal-slices}. The evolution of the sister trajectory at $J\sim 14$ is a very distinct feature. In addition, we observe at $N_{\rm max} = 10,12$ a transient ``niece trajectory'', which is pushed to higher levels $n$ with growing $N_{\rm max}$. The transitional character of the niece trajectory can be better traced and studied in fig.~\ref{fig:couplings-gif}.

\section{Removing states on leading trajectory}
\label{app:state-removal}

Beyond the main objective of this study regarding the necessity of infinite trajectories, our primal bootstrap approach offers a unique testing ground for recent analytic results constraining the spectrum of weakly-coupled theories. Specifically, Berman and Geiser recently derived Sequential Spin Constraints (SSC) and Sequential Mass Constraints (SMC) for gravitational and scalar theories \cite{Berman:2024owc}. These theorems impose strict ordering requirements on the spectrum: for example, the lightest spin-$j$ state must be lighter than the lightest spin-$(j+2)$ state. 

In the context of our ansatz in eq.~\ref{eq:primal-ansatz}, the states subject to these tightest constraints lie on the leading trajectory $j(t)=2t+2$. Since the majority of the ansatz terms saturate the SSC bounds, the associated couplings should be rigid, and removing a state from the leading trajectory should force the rest of the couplings on the leading trajectory to vanish as well.

To test this, we added an auxiliary constraint to our optimization procedure of sec.~\ref{sec:numerics-ansatz}. We explicitly enforced the vanishings of specific couplings on the leading trajectory, creating ``holes'' in the spectrum in the leading trajectory. For instance, we maximized the coupling $c_{1,4}^{++}$ subject to the additional constraint $c_{2,6}^{++}\equiv0$.

\begin{figure}[H]   
\includegraphics[width=0.49\linewidth]{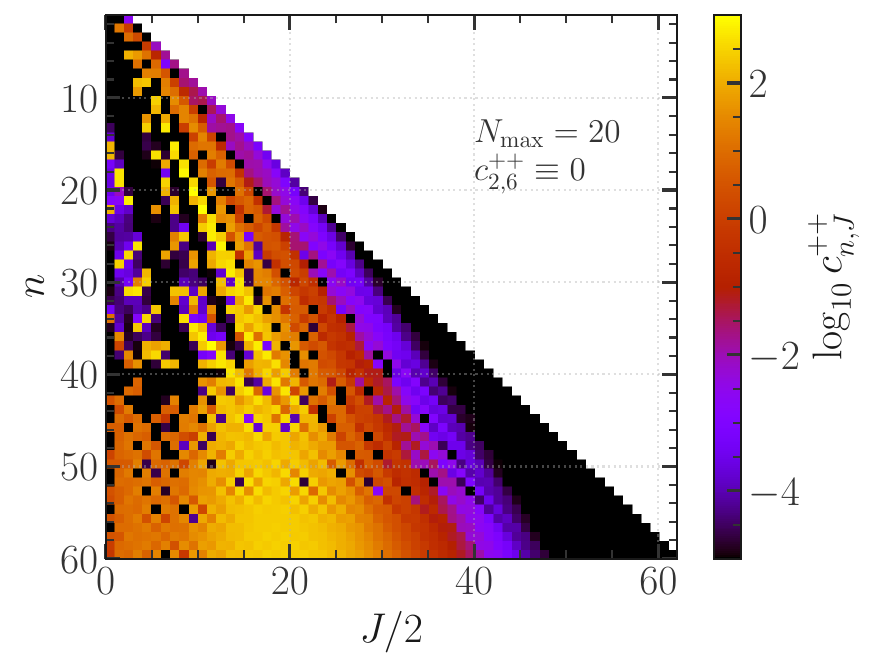}\hfill \includegraphics[width=0.49\linewidth]{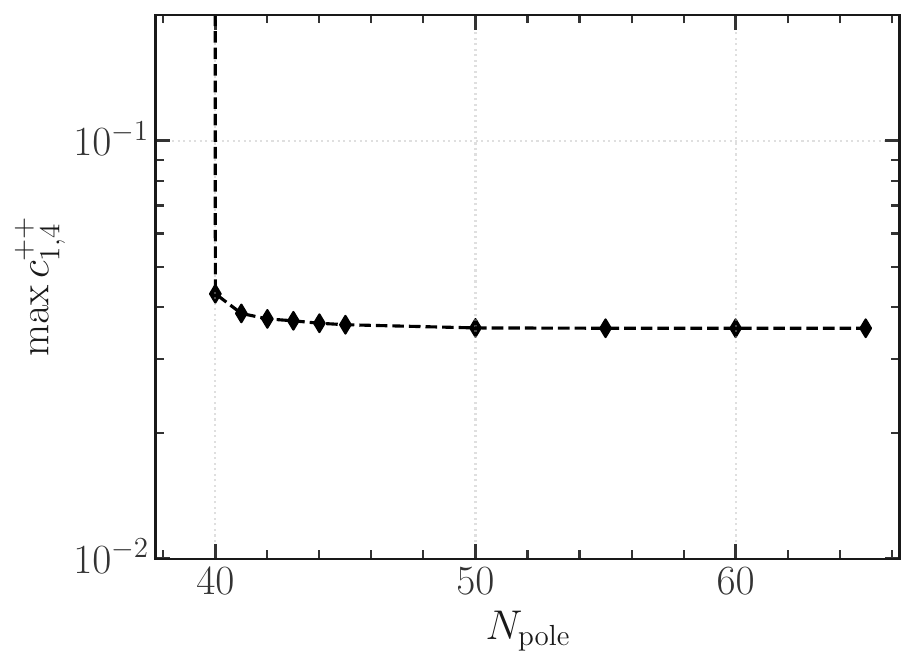}
    \caption{We illustrate with the left panel how the couplings of the extremal amplitude (maximizing $c^{++}_{1,4}$) look like once the coupling $c^{++}_{2,6}$ on the leading trajectory is forced to be identically zero. We find an admissible amplitude satisfying all positivity constraints and, as shown in the right panel, exhibiting quick convergence with the number of added positivity constraints $N_{\rm pole}$ (using all couplings $c_{n,J}^{++}$ and $c_{n,J}^{+-}$ for $n\leq N_{\rm pole}$).}
    \label{fig:remove-cnj}
\end{figure}

Surprisingly, we seem to observe that valid solutions are found with such excised leading trajectories. As illustrated in the left panel of fig.~\ref{fig:remove-cnj}, the SDPB optimization identifies a consistent amplitude configuration that satisfies positivity constraints ($c_{n,J}^{++}\geq0$ and $c_{n,J}^{+-}\geq0$ for $n\leq N_{\rm pole}$) despite the forced absence of the spin-6 state at level $n=2$. The resulting spectrum exhibits a redistribution of weights among the subleading trajectories and the remaining leading states, and maintains the overall single-trajectory-like transient structure.

We further imposed even stronger constraints, demanding that  $c_{1,4}^{++}\equiv0$,  $c_{3,8}^{++}\equiv0$, or indeed that all three of these leading couplings vanish simultaneously. In all cases, feasible solutions were found.

The right panel of fig.~\ref{fig:remove-cnj} demonstrates the stability of these solutions; the extremal value of $c_{1,4}^{++}$ (in the case where $c_{2,6}^{++}\equiv0$) shows very fast convergence as we increase the number of unitarity constraints $N_{\rm pole}$. 

Overall, these results seem in tension with the theorems of  \cite{Berman:2024owc}. A simple resolution of this tension might be the fact that $f(s|t, u)$ is fundamentally $ut$-symmetric, which escapes the assumptions of the SMC and SSC bounds. It would be very interesting to further investigate this point.

\bibliographystyle{JHEP}
\bibliography{biblio.bib}

\end{document}